\documentclass[journal]{IEEEtran}
\IEEEoverridecommandlockouts

\usepackage{cite}
\usepackage{url}
\usepackage{graphicx}
\usepackage{subfigure}
\usepackage[table,xcdraw]{xcolor}
\usepackage{adjustbox}
\usepackage{array}
\usepackage{multirow}
\usepackage{makecell}
\usepackage{balance}
\usepackage{amsmath}
\usepackage{booktabs}
\newcolumntype{L}[1]{>{\raggedright\let\newline\\\arraybackslash\hspace{0pt}}m{#1}}
\newcolumntype{C}[1]{>{\centering\let\newline\\\arraybackslash\hspace{0pt}}m{#1}}
\newcolumntype{R}[1]{>{\raggedleft\let\newline\\\arraybackslash\hspace{0pt}}m{#1}}
\usepackage{dblfloatfix}

\usepackage{blindtext}
\usepackage{booktabs} 
\usepackage{pifont}
\usepackage{array}

\usepackage{fixltx2e}  

\usepackage{adjustbox}
\usepackage{tabularx} 
\usepackage{pifont}
\usepackage{multicol}
\usepackage{here} 
\newcommand{\ie}{{\em i.e., }}
\newcommand{\eg}{{\em e.g., }}

\newcommand{\smallsim}{\smallsym{\mathrel}{\sim}}
\makeatletter
\newcommand{\smallsym}[2]{#1{\mathpalette\make@small@sym{#2}}}
\newcommand{\make@small@sym}[2]{%
	\vcenter{\hbox{$\m@th\downgrade@style#1#2$}}%
}
\newcommand{\downgrade@style}[1]{%
	\ifx#1\displaystyle\scriptstyle\else
	\ifx#1\textstyle\scriptstyle\else
	\scriptscriptstyle
	\fi\fi
}
\makeatother

\ifCLASSINFOpdf
\else
\fi
\begin{document}

\title{Comparing Broadband ISP Performance\\ using Big Data from M-Lab}

\author{Xiaohong~Deng,
              Yun~Feng,
              Thanchanok~Sutjarittham,
              Hassan~Habibi~Gharakheili,
              Blanca~Gallego,
			  and~Vijay~Sivaraman
	    \IEEEcompsocitemizethanks{
	    	\IEEEcompsocthanksitem X.~Deng was with the School of Electrical Engineering and Telecommunications, University of New South Wales, Sydney, NSW 2052, Australia (e-mail: dxhbupt@gmail.com).	    	   	
	    	\IEEEcompsocthanksitem T.~Sutjarittham, H.~Habibi~Gharakheili, and V. Sivaraman are with the School of Electrical Engineering and Telecommunications, University of New South Wales, Sydney, NSW 2052, Australia (e-mails: t.sutjarittham@unsw.edu.au, h.habibi@unsw.edu.au, vijay@unsw.edu.au).
	    	\IEEEcompsocthanksitem {Y.~Feng is with Shanghai Huawei Technologies, Pudong, China, ZIP 201206 (e-mail: fengyun15@huawei.com)}.	    	
	    	\IEEEcompsocthanksitem B.~Gallego is with the Centre for Big Data Research in Health, University of New South Wales, Sydney, NSW 2052, Australia (e-mail: b.gallego@unsw.edu.au).
            \IEEEcompsocthanksitem This submission is an extended and improved version of our paper presented at the ITNAC 2015 conference \cite{itnac2015}	    	
	    	
	}
}

\markboth{}%
{Deng \MakeLowercase{\textit{et al.}}: Bare Demo of IEEEtran.cls for IEEE Journals}

\maketitle

\begin{abstract}
Comparing ISPs on broadband speed is challenging, since measurements can vary due to subscriber attributes such as operation system and test conditions such as access capacity, server distance, TCP window size, time-of-day, and network segment size. In this paper, we draw inspiration from observational studies in medicine, which face a similar challenge in comparing the effect of treatments on patients with diverse characteristics, and have successfully tackled this using ``causal inference'' techniques for {\em post facto} analysis of medical records. Our first contribution is to develop a tool to pre-process and visualize the millions of data points in M-Lab at various time- and space-granularities to get preliminary insights on factors affecting broadband performance. Next, we analyze 24 months of data pertaining to twelve ISPs across three countries, and demonstrate that there is observational bias in the data due to disparities amongst ISPs in their attribute distributions. For our third contribution, we apply a multi-variate matching method to identify suitable cohorts that can be compared without bias, which reveals that ISPs are closer in performance than thought before. Our final contribution is to refine our model by developing a method for estimating speed-tier and re-apply matching for comparison of ISP performance. Our results challenge conventional rankings of ISPs, and pave the way towards data-driven approaches for unbiased comparisons of ISPs world-wide. 
\end{abstract}


\section{Introduction} \label{sec:intro}
This paper asks the question: how should we compare Internet Service Providers (ISPs) in terms of the broadband speeds they provide to consumers? (aspects such as pricing plans, quotas, and reliability are not considered in this paper). On the face of it, determining the answer may seem simple: a subscriber's speed can be measured directly (say via a speed-test tool or an adaptive bit-rate video stream), allowing ISPs to be compared based on the average (or median) measured speed across their subscriber base. However, this approach has deep conceptual problems: an ISP-A who has many subscribers in remote areas served by low-capacity (wired or wireless) infrastructure will compare poorly to an ISP-B whose subscribers are predominantly city dwellers connected by fiber; yet, it could well be that ISP-A can provide higher speeds than ISP-B to every subscriber covered by ISP-B! The comparison bias illustrated above arising from disparity in access capacity is but one example of many potential confounding factors, such as latency to content servers, host and server TCP window size settings, maximum segment size in the network, and time-of-day, that directly bias measurement test results. Observational studies therefore need to understand and correct for such biases to ensure that the comparisons are fair.

In this study we draw inspiration from the field of medicine, which has grappled for decades with appropriate methods to compare new drugs/treatments. ``Patients'' (in our case broadband subscribers) with ``attributes'' such as gender, age, medical conditions, and prior medications (in our case access-capacity, server-latency, host-settings, etc.) are given ``treatments'' (in our case ISPs), and their efficacy needs to be compared. The dilemma is that any given patient can only be measured taking treatment-A or treatment-B, but not both at the same time; similarly, a subscriber in our case can only be observed when connected to one ISP, so the ``ground truth'' of that customer’s broadband performance served by other ISPs is never observed. To overcome this issue, the gold standard for medical treatment comparisons is a ``randomized control trial'' (RCT), wherein each patient in the cohort is {\it randomly} assigned to one of the multiple treatments (one of which could be a placebo). The randomization is crucial here, in the expectation that known as well as unknown attributes that could confound the experiment outcome get evenly distributed across the groups being compared so that statistically meaningful inferences can be drawn.

Alas, ``randomized'' assignment of ISPs to subscribers is not a viable option in the real world, so we have to instead rely on ``observational'' studies that analyze performance data given {\em a priori} assignment of ISPs to subscribers. Fortunately for us, techniques for observational studies are maturing rapidly, particularly in medicine where analyzing big data from electronic health records is much cheaper than running controlled clinical trials, and can yield valuable insights on the causal relationship between patient attributes and treatment outcomes. In this work we have collaborated closely with a medical informatics specialist to apply ``causal inference'' techniques to analyzing ISP performance data -- unlike a classic supervised learning problem, causal inference works by estimating how things might look in different conditions, thereby differentiating the influence of A versus B, instead of trying to predicting the outcome. We apply this method to the wealth of broadband data available from the open M-Lab platform, which holds over 40 million measurement results world-wide for the year 2016. Though no data-driven approach can guarantee that causal relationships are deduced correctly, as there could be unknown attributes that affect outcome (the ``unknown unknowns'', to use the Rumsfeld phrase), we believe that the M-Lab data set captures most, if not all, of the important attributes that are likely to affect the speed measurements.

Our objective in this paper is to apply emerging data-analysis techniques to the big-data from M-Lab to get new insights into ISP broadband performance comparison. Our first contribution is somewhat incidental - we develop a tool that allows researchers to easily and quickly process and depict M-Lab data to visualize performance metrics (speed, latency, loss, congestion) at various spatial (per-house, per-ISP, per-country) and temporal (hourly, monthly, yearly) granularities. Our second contribution applies our tool to over 17 million data samples taken in 2015 and 2016 spanning 12 ISPs in 3 countries, to identify the relative impact of various attributes (access speed-tier, host settings, server distance, etc.) on broadband performance. We reveal, both visually and analytically, that dominant attributes can vary significantly across ISPs, corroborating our earlier assertion that subscriber cohorts have disparate characteristics and ISP comparisons are therefore riddled with bias. 
Our third contribution is to apply a causal inference technique, called multi-variate matching, to filter the data sets by identifying cohorts with similar attributes across ISPs. Our final contribution is to refine our method to make M-Lab data more useful by mapping measurements with households and estimate their speed-tier, so that meaningful performance comparisons across households can be conducted.
Our results indicate that the ISPs are closer in speed performance than previously thought, and their relative ranking can be quite different to what the raw aggregates indicate.

The rest of this paper is organized as follows: \S\ref{sec:prior-work} recaps prior work in broadband performance, and gives relevant background on causal inference techniques. In \S\ref{sec:mlab} we describe our measurement data set, the attributes it contains, and preliminary insights gleaned from our visualization tool. The attribute distributions and underlying biases are discussed in \S\ref{sec:data-attribs}, while in \S\ref{sec:matching} we apply multi-variate matching to reduce bias and compare ISPs in a fair manner. \S\ref{sec:refinedMatching} presents our systematic approach  to estimate household access capacity from M-Lab data. The paper is concluded in \S\ref{sec:concl} with pointers to future work.


\section{Background and Prior Work} \label{sec:prior-work}

\subsection{Broadband Measurement and Reporting} \label{sec:prior-work-measurement}
Measuring and ranking broadband ISPs has been ongoing (and contentious) for several years -- Netflix publishes a monthly ISP speed index \cite{Netflix} that ranks ISPs based on their measured prime time Netflix performance, and Youtube graphs for each ISP the break-down of video streaming quality (low vs standard vs high definition) by time-of-day averaged over a 30-day period \cite{Youtube}. While these large content providers undoubtedly have a wealth of measurement data, these are specific to their services, and neither their data nor their precise comparison methods are available in the public domain (to be fair Google does outline a methodology on its video quality report page, but it fails to mention important elements such as whether it only considers video streams of a certain minimum duration, whether a house that watches more video streams contributes more to the aggregate rating, and how it accounts for various factors such as browser type, server latency, etc. that vary across subscribers and can affect the measurement outcome). Governments are also under increasing pressure to compile consumer reports on broadband performance -- for example the FCC in the US \cite{FCC} directs consumers to various speed test tools to make their own assessment, and the ACCC in Australia \cite{ACCC} is running a pilot program to instrument volunteers' homes with hardware probes to measure their connection speeds.
Additionally, various national regulators in Europe employ their own methods of measuring broadband speed and publish white papers as surveyed in \cite{EUAnalysis} -- for example, the ofcom in the UK uses a hardware measurement unit (developed by SamKnows), while several other national regulators such as in Italy, Austria, Germany, Portugal, Slovenia use specialized software solutions (developed in-house) while the regulator in Greece adopted M-Lab's NDT tool.

While there is a commendable amount of effort being expended on collecting data, via either passive measurement of video traffic or active probing using hardware devices (we refer the reader to a recent survey \cite{survey15} that gives an overview of measurement platforms and standardization efforts), less effort has been expended on a systematic analysis of the collected data. This matters, because early works such as \cite{MIT10} have demonstrated that broadband speed measurements can exhibit high variability, and these differences arise from a complex set of factors including test methodology and test conditions, including home networks and end users' computers, that make it very challenging to attribute performance bottlenecks to the constituent parts, specifically the ISP network. While their work acknowledges that broadband benchmarking needs to look beyond statistical averages to identify factors exogenous to the ISP, they do not offer any specific approaches for doing so. 
NANO \cite{NANO} developed a system that infers the effect of ISPs policy on a service performance -- it also compares the service performance across multiple ISPs. NANO establishes a causal relationship between an ISP and its observed performance by adjusting confounding factors such as client-based confounder, network-based confounder and time-based confounder. But, it does not consider the TCP throughput performance comparison across ISPs.
NetDiff \cite{NetDiff} designed a system that offers a fair performance comparison of ISP networks by taking into account the size and geographic spread of each ISP. It helps customers determine which ISP offers the best performance according to their specific workload. But their work considers only one confounding factor.
A separate body of work \cite{sigcomm02,imc04,sigmetrics07} explores model-driven and data-driven methods to estimate or predict end-to-end available bandwidth; however, they operate at short time-scale, their data-sets are small, and their focus is not specific to broadband networks. We believe our work is among the first to combine causal inference techniques for observational studies with the big data openly available from the M-Lab measurement platform to attempt a fair comparison of ISP broadband performance. 

\subsection{Causal Inference Analysis} \label{sec:prior-work-causal-inference}
As mentioned earlier, the gold standard for comparisons is a randomized control trial, which is not feasible in our case. We therefore have to use observational data with {\em a priori} assignments of ISPs to subscribers, and use causal inference methods \cite{CIOVERVIEW,CausalPotential,ci1} that can control for differences in the covariate distributions between the groups being compared so as to minimize confounding. One of the most popular methods is ``matching'' \cite{matchreview}, which selects subsets of observations in one group (the {\em treatment group}) for comparison with observations having similar covariate distributions in the comparator group (the {\em control group}) -- balancing the distribution of covariates in the two groups gives the effect of a randomized experiment. Matching has been used extensively in epidemiological, social, and economic research studies, and has been proven to reduce confounding bias very effectively. The most common approaches to perform matching are propensity score matching, multivariate matching based on Mahalanobis distance, and more recently, genetic matching algorithms. In this paper we chose multivariate matching, which is easier to tune by interpreting results when the number of attributes is not too large, and is well supported in R \cite{Matching11}.

Once the covariates of the observations have been matched between the groups, the difference in outcome is averaged to estimate the {\em average treatment effect (ATE)} -- in medicine, this could quantify the average effect of a drug compared to a placebo, while in our case it estimates the average difference in download speed between the two ISPs being compared. Certain pre-conditions are needed for our approach: we assume that the group assignment (\ie choice of ISP for subscriber) has been made independent of the outcome, conditional on the observed covariates; that the baseline covariates, although measured {\em post facto}, are not affected by the treatment (ISP); and that there are sufficient observations for which the probability of assignment is bounded away from zero and one. In simple words, this states that households (patients) did not make ISP (treatment) choice based on known outcomes or attributes, and there are a reasonable number of samples from the two groups being compared that have similar covariate distributions (our results in \S\ref{sec:matching} will capture this via p-values).


\begin{figure}[!t]
	\centering
	\includegraphics[width=0.5\textwidth]{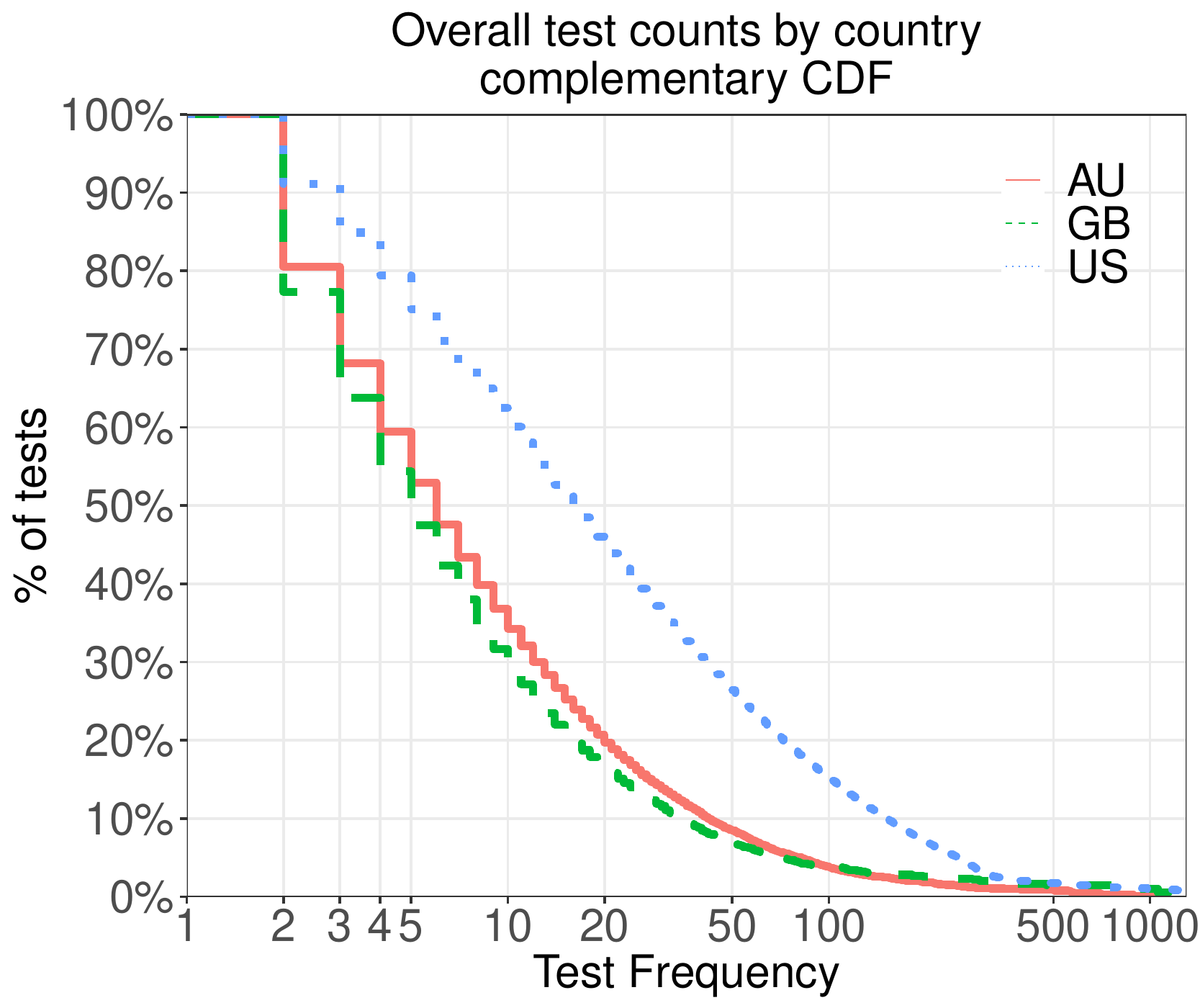}
	\vspace{-6mm}
	\caption{CCDF of household test counts in AU, UK, and US.}
	\vspace{-4mm}
	\label{fig:3testcounts}
\end{figure}

\section{Data-Set Selection, Attributes, Visualization} \label{sec:mlab}
In this section we briefly introduce M-Lab, its measurement tools and data repositories. We then describe the data we have selected and pre-processed, the attributes we have extracted, and the visualization tool we have built.

M-Lab \cite{M-Lab} was founded in 2008 as a consortium of research, industry, and public-interest partners, backed by Google, to create a global open measurement platform and data repository that researchers can use for deeper studies of Internet performance. M-Lab has built a platform on which test servers are well distributed across continents and ISPs, and any interested party can design, implement, and deploy new Internet measurement tools under an open license. 
This gives a significant advantage to M-Lab over other platforms such as PerfSONAR \cite{perfSONAR} -- M-Lab provides a much larger collection data generated by tens of millions of tests from clients connected to hundreds of ISPs across the globe every year.
All data collected in the M-Lab platform are open access (as opposed to commercial platforms such as Ookla \cite{Ookla}), and available either in raw format as file archives over Google cloud storage, or in SQL-friendly parsed format accessible using BigQuery. In terms of diversity, the M-Lab covers a more diverse range of users compared to hardware-based platforms such as SamKnows \cite{Samknows} and BISMark \cite{BISmark} --  deploying a hardware-based measurement at users' premises is constrained by distribution of devices, and thus limited to selected populations.

\begin{figure*}[t!]
	\begin{center}
		\mbox{
			\subfigure[AU.]{
				{\includegraphics[width=0.3\textwidth]{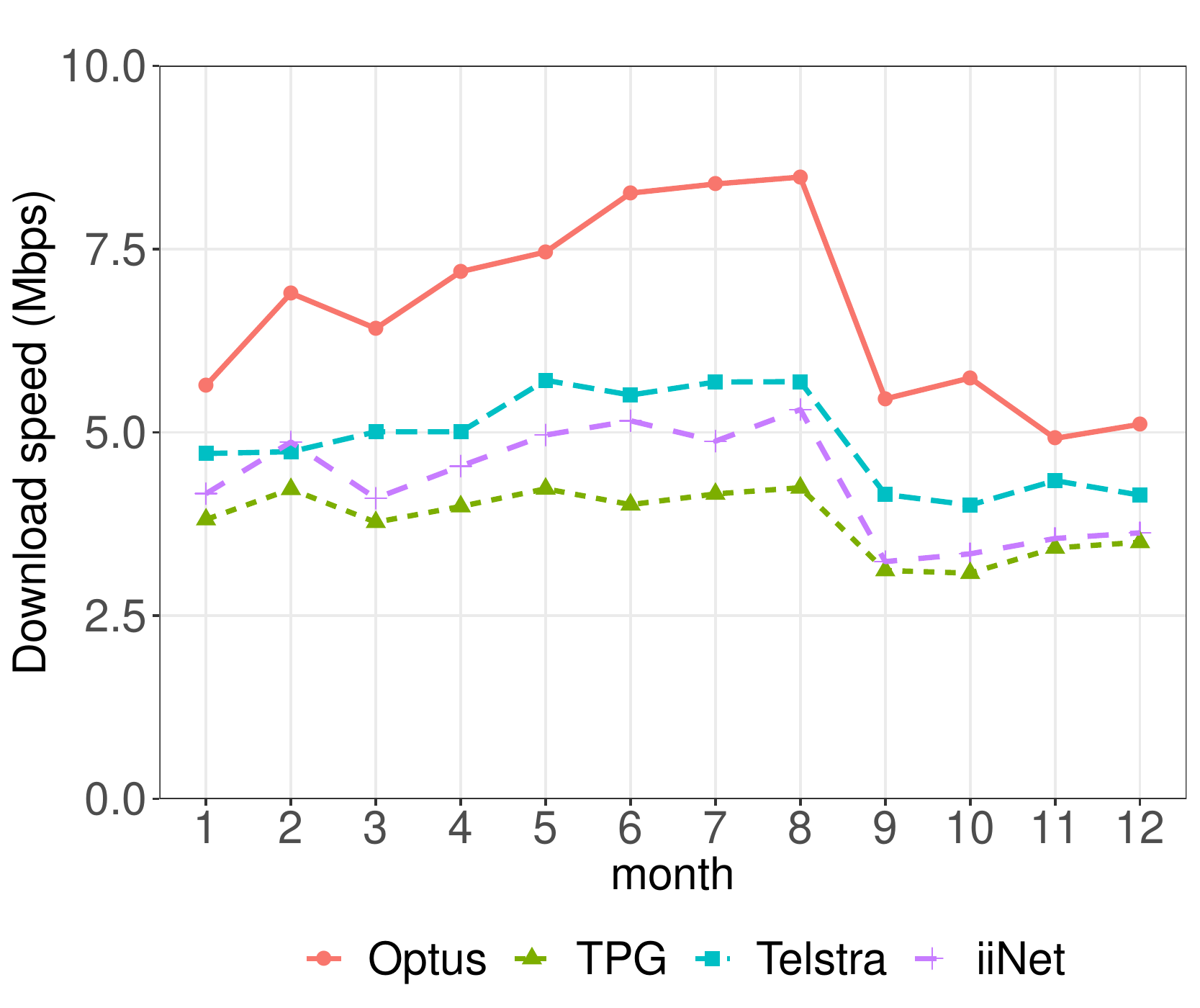}}\quad
				\label{fig:3monthlymedianAUhouse}
			}
			\hspace{-2mm}
			\subfigure[UK.]{
				{\includegraphics[width=0.3\textwidth]{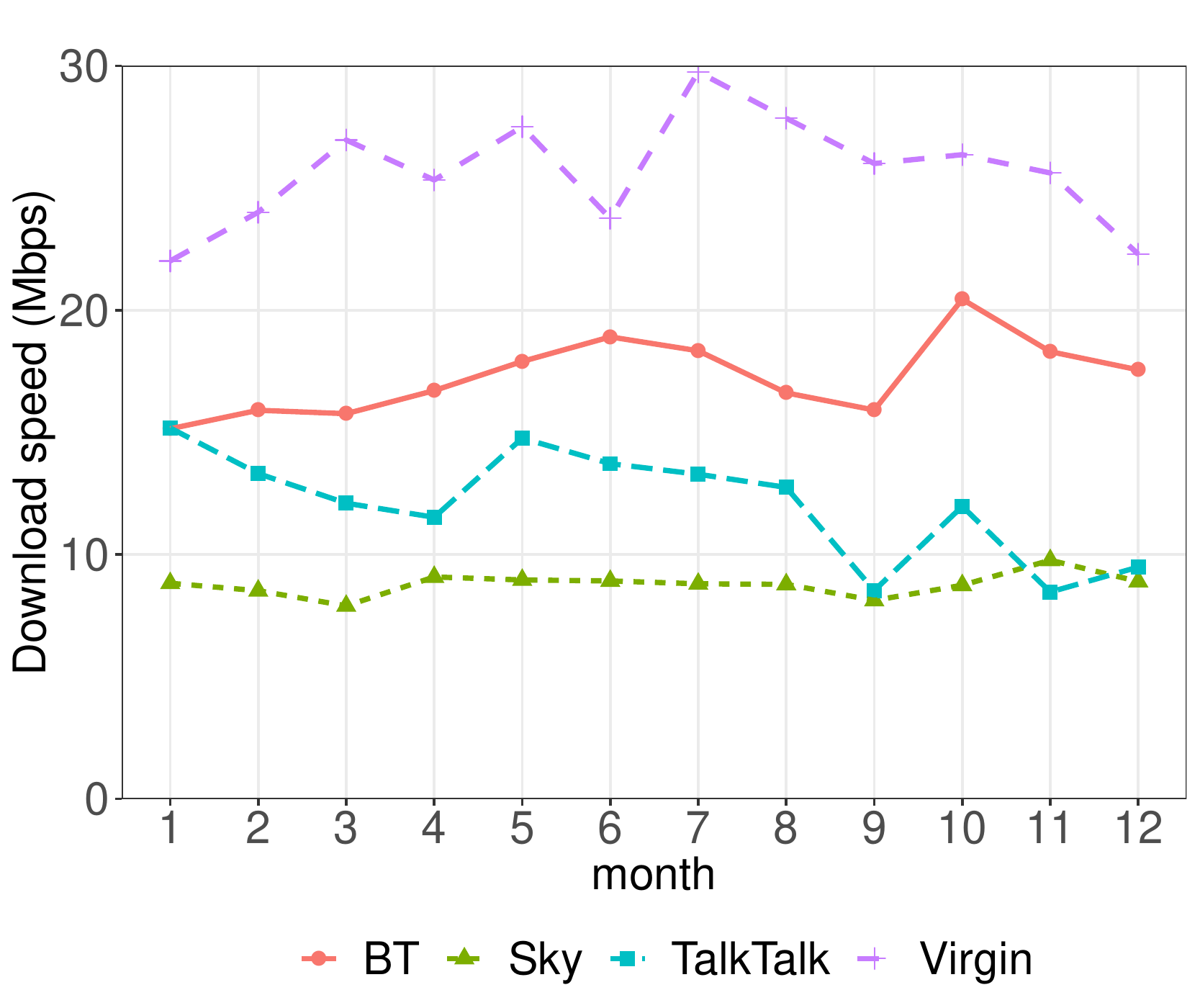}}\quad
				\label{fig:3monthlymedianUKhouse}
			}
			\hspace{-2mm}
			\subfigure[US.]{
				{\includegraphics[width=0.3\textwidth]{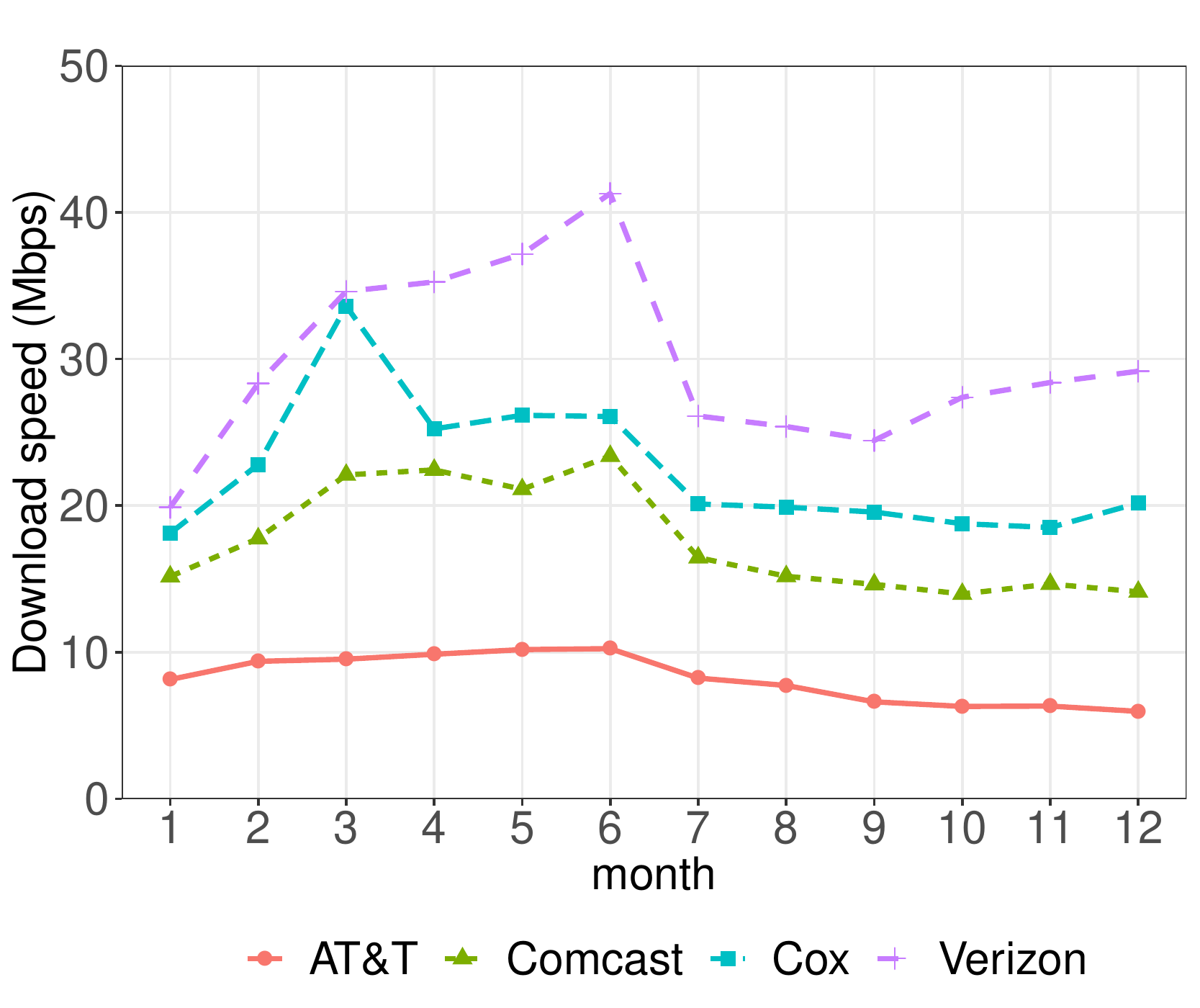}}\quad
				\label{fig:3monthlymedianUShouse}
			}
		}
		\vspace{-2mm}
		\caption{Monthly median download speeds: (a) Australia, (b) UK, (c) US.}
		\vspace{-5mm}
		\label{fig:3monthlymedianhouse}
	\end{center}
\end{figure*}

\subsection{Data Set Selection and Pre-Processing} \label{sec:mlab-data}

In this paper we use the data collected by the Network Diagnostic Test (NDT) tool, since it has by far the largest number of speed test samples (over 40 million for the year 2016), and captures a rich set of attributes for each test (discussed later in this section). In order to evaluate the generality of our methods, we apply them to data from three countries: Australia (AU), the United Kingdom (UK), and the United States (US). We select four of the largest ISPs from each country for comparison: Telstra, Optus, iiNet, and TPG from AU; BT, Virgin, Sky and TalkTalk from the UK; and Comcast, Verizon, AT\&T, and Cox from the US. For these ISPs, we analyze the NDT speed test measurements taken from two years (2015 and 2016), comprising $1.3$m samples for AU, $1.4$m for UK, and $14.5$m for US -- the latter is an order of magnitude larger since Google searches in the US got linked with NDT as of July 2016.

{\bf Determining household speed:} Our first objective is to set a baseline for ISP speed comparison by computing their mean/median values. However, we found in the NDT data-set that some IP addresses were conducting many more tests than others (for convenience of exposition we will henceforth refer to each unique IP address as a ``household''). Our data set was found to have around $565$k, $464$k, and $2.81$m households for AU, UK, and US respectively, indicating that the average household contributes only $2$-$4$ samples each month. There is however a significant skew in monthly test frequency amongst households, as shown in the complementary CDF (CCDF) of Fig.~\ref{fig:3testcounts} - in AU and UK for example, the bottom 50\% of households contribute 5 or fewer samples, while the top 10\% contribute 50 or more samples each. We eliminate this bias by aggregating (averaging) the results to obtain a single value per household per month, and plot the resulting month-by-month median download speed across households in Fig.~\ref{fig:3monthlymedianhouse}. The rankings shown in the figure are broadly consistent with the ones published by Netflix -- like us, the Netflix ISP speed index ranks Optus, Virgin, and Verizon as highest respectively in AU, UK, and US for most months of 2016. There is no reference point to how Netflix computes its ISP indexing. But because this way it generates similar ranking as Netflix index, so we believe it's a fair baseline and will use it as a naive baseline for our discussions and comparisons against other methods developed later in this paper. Therefore, later on, when we refer to naive arithmetic average, we mean ISP performance aggregated by household -- each house only has one vote to the ISP average.

\begin{table}[t!]
	\caption{Number of Sampled NDT measurements from each ISP}
	\label{tab:remaining-samples}
	\begin{tabular}{|c|c|c|c|c|}
		\hline
		year & ISP      & raw test count & annual test count $>$ 20 & used   \\ \hline
		
		2015 & Telstra  & 117,019        & 11,180                     & 9.6\%  \\ \hline
		
		2015 & Optus    & 46,138         & 8,232                      & 17.8\% \\ \hline
		
		2015 & iiNet    & 42,917         & 5,144                      & 12.0\% \\ \hline
		
		2015 & TPG      & 52,186         & 14,928                     & 28.6\% \\ \hline
		
		2015 & BT       & 238,134        & 13,844                     & 5.8\%  \\ \hline
		
		2015 & Virgin   & 205,149        & 54,371                     & 26.5\% \\ \hline
		
		2015 & Sky      & 235,271        & 27,623                     & 11.7\% \\ \hline
		
		2015 & TalkTalk & 7,450          & 652                        & 8.8\%  \\ \hline
		
		2015 & AT\&T    & 460,482        & 148,512                    & 32.3\% \\ \hline
		
		2015 & Cox      & 215,499        & 56,282                     & 26.1\% \\ \hline
		
		2015 & Verizon  & 291,421        & 88,010                     & 30.2\% \\ \hline
		
		2015 & Comcast  & 769,728        & 216,252                    & 28.1\% \\ \hline
		
		2016 & Telstra  & 478,469        & 65,839                     & 13.8\% \\ \hline
		
		2016 & Optus    & 161,303        & 26,421                     & 16.4\% \\ \hline
		
		2016 & iiNet    & 199,764        & 20,525                     & 10.3\% \\ \hline
		
		2016 & TPG      & 219,547        & 53,879                     & 24.5\% \\ \hline
		
		2016 & BT       & 126,013        & 10,890                     & 8.6\%  \\ \hline
		
		2016 & Virgin   & 71,947         & 20,405                     & 28.4\% \\ \hline
		
		2016 & Sky      & 95,964         & 10,292                     & 10.7\% \\ \hline
		
		2016 & TalkTalk & 17,543         & 3,666                      & 20.9\% \\ \hline
		
		2016 & AT\&T    & 2,841,976      & 1,280,814                  & 45.1\% \\ \hline
		
		2016 & Cox      & 1,391,983      & 580,976                    & 41.7\% \\ \hline
		
		2016 & Verizon  & 1,333,403      & 537,319                    & 40.3\% \\ \hline
		
		2016 & comcast  & 5,442,720      & 2,164,122                  & 39.8\% \\ \hline
		
	\end{tabular}
\end{table}


{\bf Determining access speed-tier:} The download speed for a household will be limited by the capacity of its access link, which in turn is dictated by physical attributes such as medium (fiber, copper, wireless) and distance from the local exchange. It may further be constrained if the subscriber has chosen a plan with lower advertised speed. We term this maximum possible speed available to the household as its ``access speed-tier''. As we will see in the next section, this attribute is important when comparing ISPs, but is not explicitly present in the data since M-Lab is not privy to advertised speeds and subscriber plans. We, therefore, have to infer a household's access speed-tier from the measured data. We take an approximate approach of using the largest value of measured speed as the access speed-tier for that household, provided: (a) the household has conducted a minimum threshold number of tests, and (b) at least one test was conducted during off-peak hours (\ie outside of 7pm-11pm local time). Filtering by higher test-count threshold will estimate access speed-tier more accurately, but reduces the data-set by eliminating households that conduct fewer tests (see CCDF of test-counts in Fig.~\ref{fig:3testcounts}). We chose threshold of $20$ for AU and UK, and $50$ for US, so as to get reasonable confidence in our estimates of access speed-tier. 
As we can see in Table~\ref{tab:remaining-samples}, the test-count thresholds we choose are able to retain 10-30\%, 6-27\% and 26-33\% samples for AU, UK and US respectively in 2015's data, and 10-24\%, 9-30\% and 40-45\% samples for AU, UK and the US respectively in 2016's data. As the magnitude of the number of samples increased in the US since July of 2016, we believe, having a test-count threshold to select measurement samples will not be a constraint for applying our method on future's data.

%
%
%

\begin{figure}[t!]
	\centering
	\includegraphics[width=0.5\textwidth]{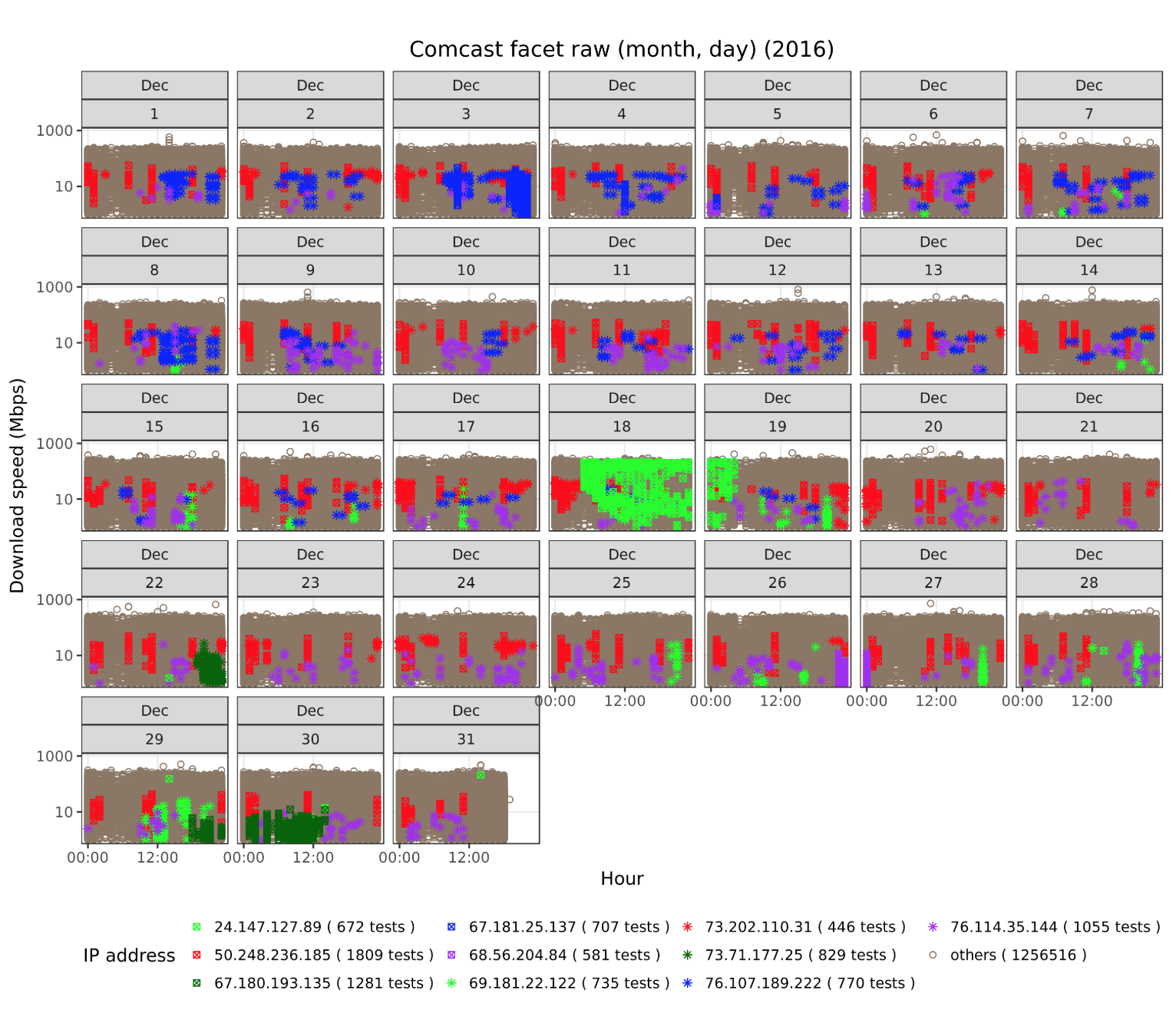}
	\vspace{-4mm}
	\caption{Comcast speed-tests faceted by day-of-month for December 2016.}
	\vspace{-4mm}
	\label{fig:3facetplotUSdaily}
\end{figure}

\subsection{Attribute Selection} \label{sec:mlab-attrib}
The NDT speed-test client connects to its nearest NDT server for the speed test. The server records the client information (IP address, geographic location, OS, client version), network attributes (RTT, MSS), server side stats (TCP buffers, maximum congestion window sizes, and other web100 variables \cite{RFC4898} for TCP tuning), and detailed run-time measurements (speed, loss, congestion signal counts). While nearly $50$ attributes are included, we found that many were sparsely recorded, due to different NDT client versions embedded in different applications on different operating systems choosing to record a different subset of attributes.

The TCP window size of the client is expected to affect measured speeds. There are a number of Web100 performance statistics related to the TCP window\cite{RFC4898}: the TCP receiver's announced maximum window size (\textit{Rwnd}) which is determined by the available head room in the \textit{RecieveBuf}, and \textit{Average Receiver Window} is generally close to \textit{Rwnd}. We use the maximum window size (\textit{Rwnd}) as it is solely dependent on client conditions, whereas \textit{RecieveBuf} can also reflect network conditions.
The {\bf distance} between client and server is recorded in the form of minimum round-trip-time (min-RTT) over the duration of each test, as is the {\bf maximum segment size (MSS)}. The {\bf client region} is recorded, which allows conversion to local {\bf time-of-day} to determine if the test is done during peak or off-peak times. The {\bf Operating System and version} attribute can tell us about the default host settings (\eg TCP auto-tuning, Nagle option) that affect speed performance. The broadband {\bf ISP} for the client is deduced based on a Whois lookup of the client IP address, and the {\bf speed-tier} for the household computed as explained earlier. 
The {\bf Client Limited Time} is the percentage of time a test was constrained by the client itself. This attribute is inherently correlated with speed-tier -- we found that higher speed-tier clients are more likely to be constrained by client-side limitations. Since this attribute is not directly associated with the speed, we do not process it when we quantify the correlation between attributes and speeds but only use it for applying the Causal Inference model later on. The impact of these attributes, reflective of test conditions, will be discussed in the next section.

\begin{figure*}[t!]
	\begin{center}
		\mbox{
			\subfigure[AU.]{ 
				{\includegraphics[width=0.30\textwidth]{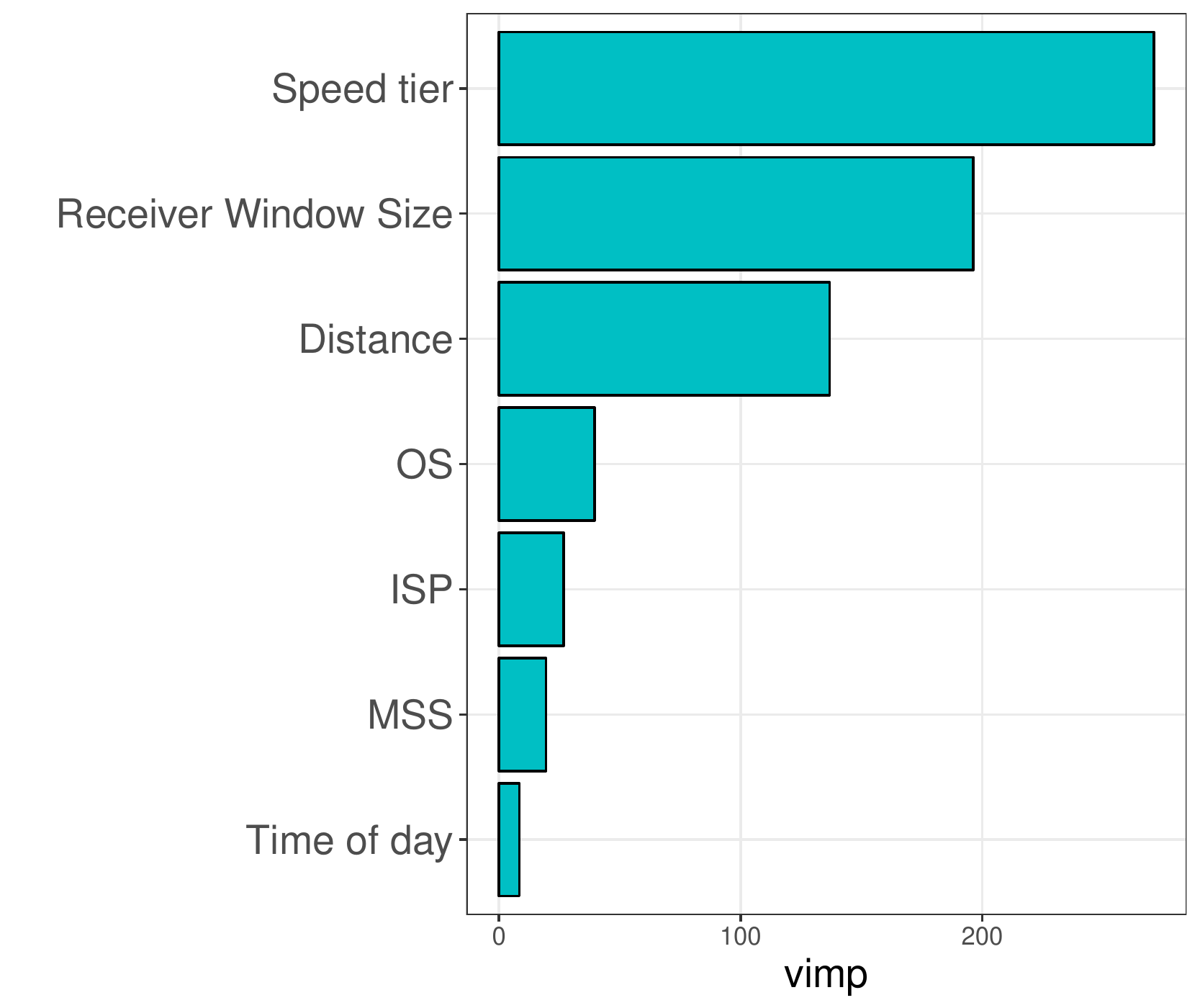}}\quad
				\label{fig:3vimpAU}
			}
			\hspace{-8mm}
			\subfigure[UK.]{ 
				{\includegraphics[width=0.30\textwidth]{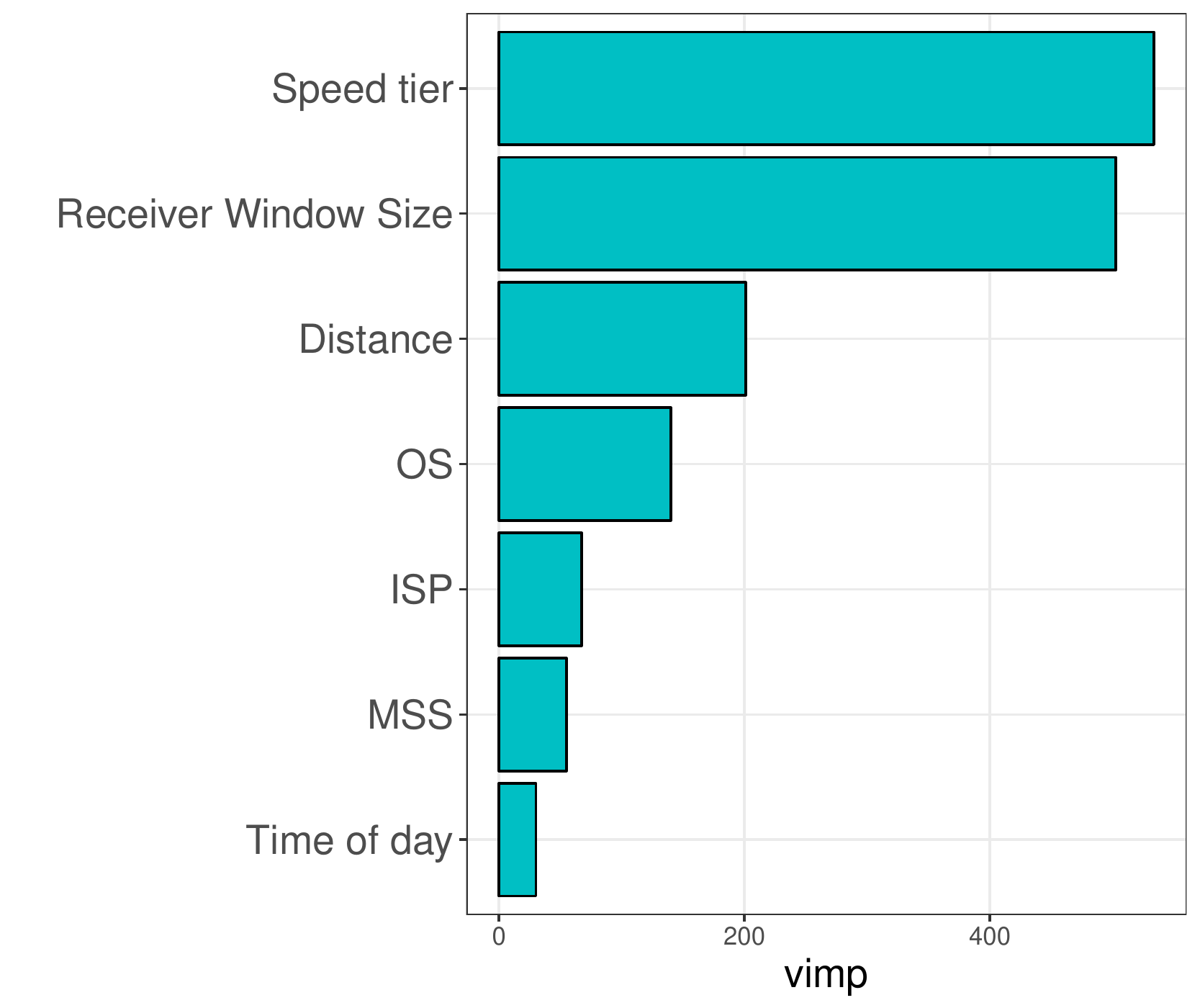}}\quad
				\label{fig:3vimpUK}
			}
			\hspace{-8mm}
			\subfigure[US.]{ 
				{\includegraphics[width=0.30\textwidth]{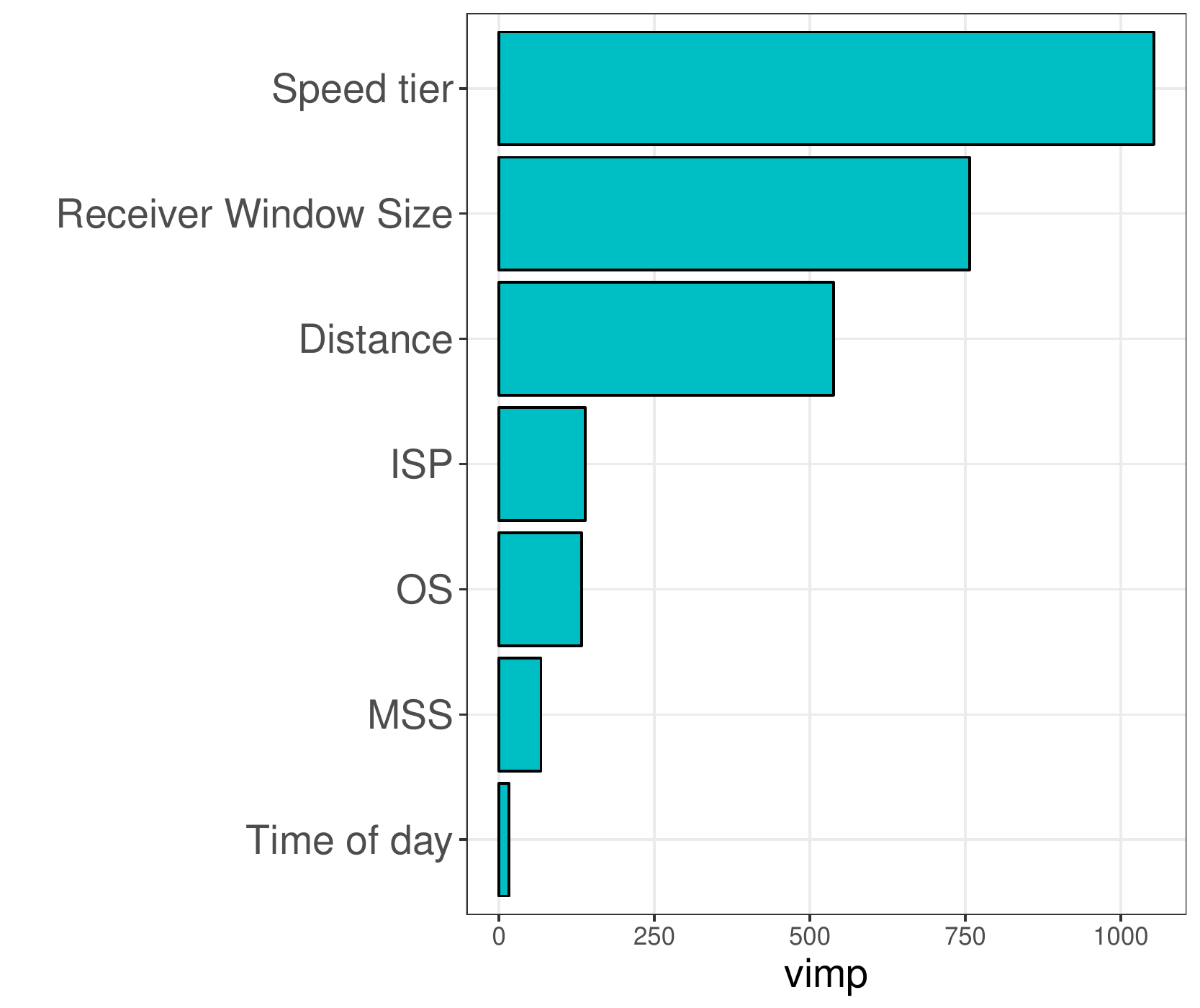}}\quad
				\label{fig:3vimpUS}
			}
		}
		\vspace{-3mm}
		\caption{Attribute importance computed from Random Forest for (a) AU, (b) UK, and (c) US.}
		\vspace{-2mm}
		\label{fig:4vimp}
	\end{center}
\end{figure*}

\begin{figure*}[t!]
	\begin{center}
		\mbox{
			\subfigure[Optus.]{ 
				{\includegraphics[width=0.43\textwidth,height=0.27\textwidth]{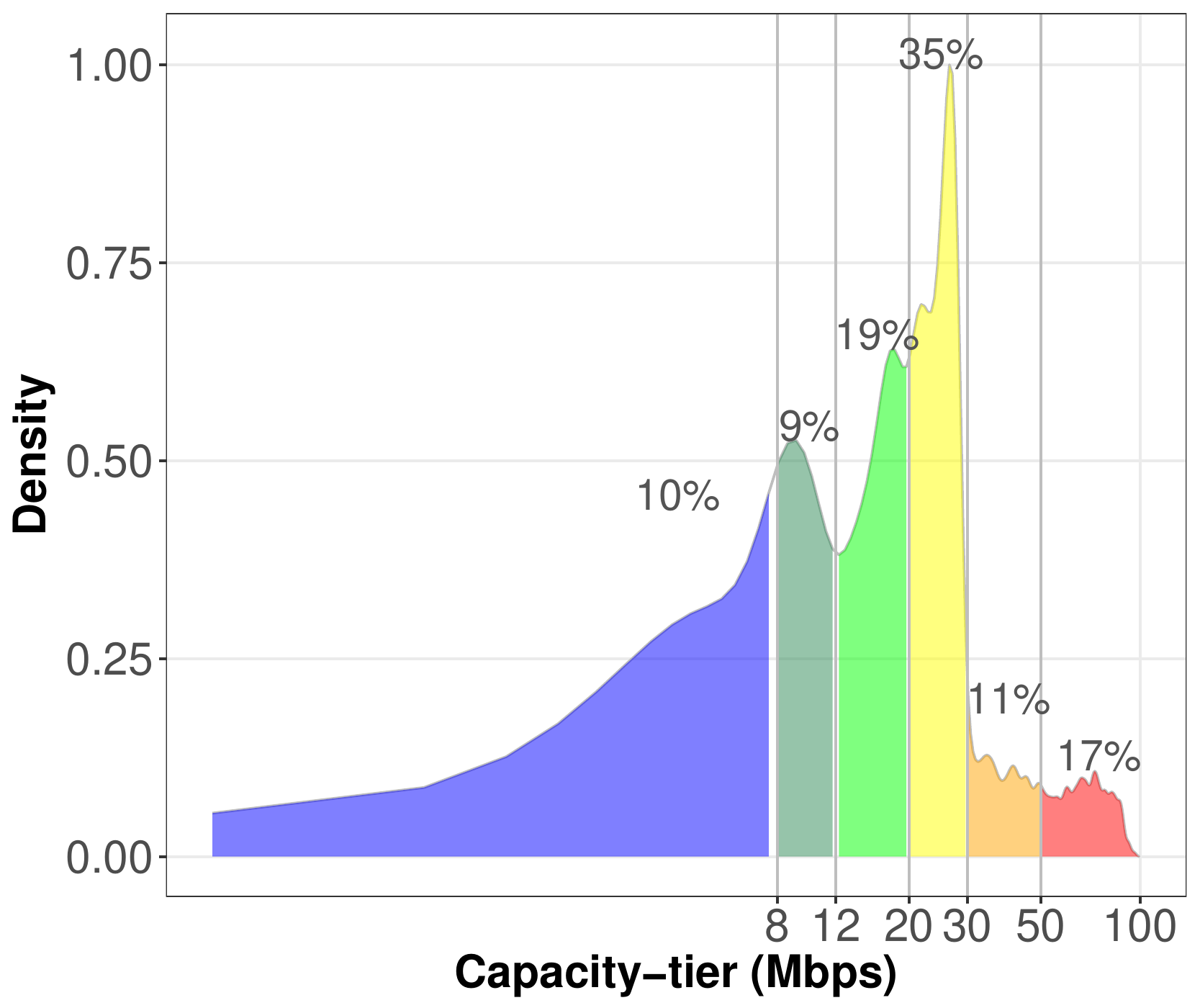}}\quad
				\label{fig:4speedtierOptus}
			}
			\hspace{-4mm}
			\subfigure[Telstra.]{ 
				{\includegraphics[width=0.43\textwidth,height=0.27\textwidth]{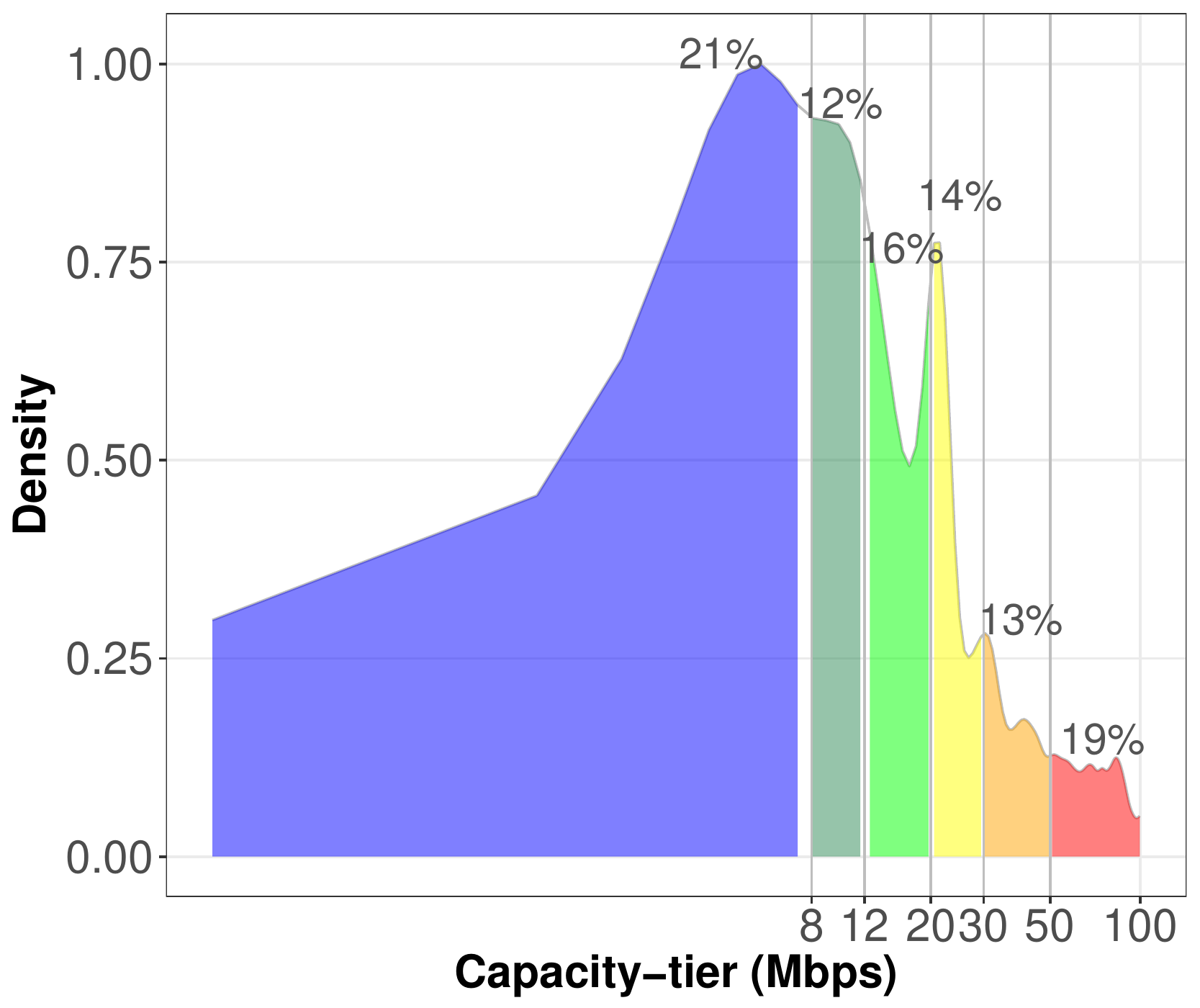}}\quad
				\label{fig:4ostypeTelstra}
			}
		}
		\vspace{-2mm}
		\caption{Comparing access speed-tier for (a) Optus, and (b) Telstra, in AU.}
		\vspace{-5mm}
		\label{fig:4speedtierAU}
	\end{center}
\end{figure*}

\begin{figure}[t!]
	\centering
	\includegraphics[width=0.43\textwidth,height=0.27\textwidth]{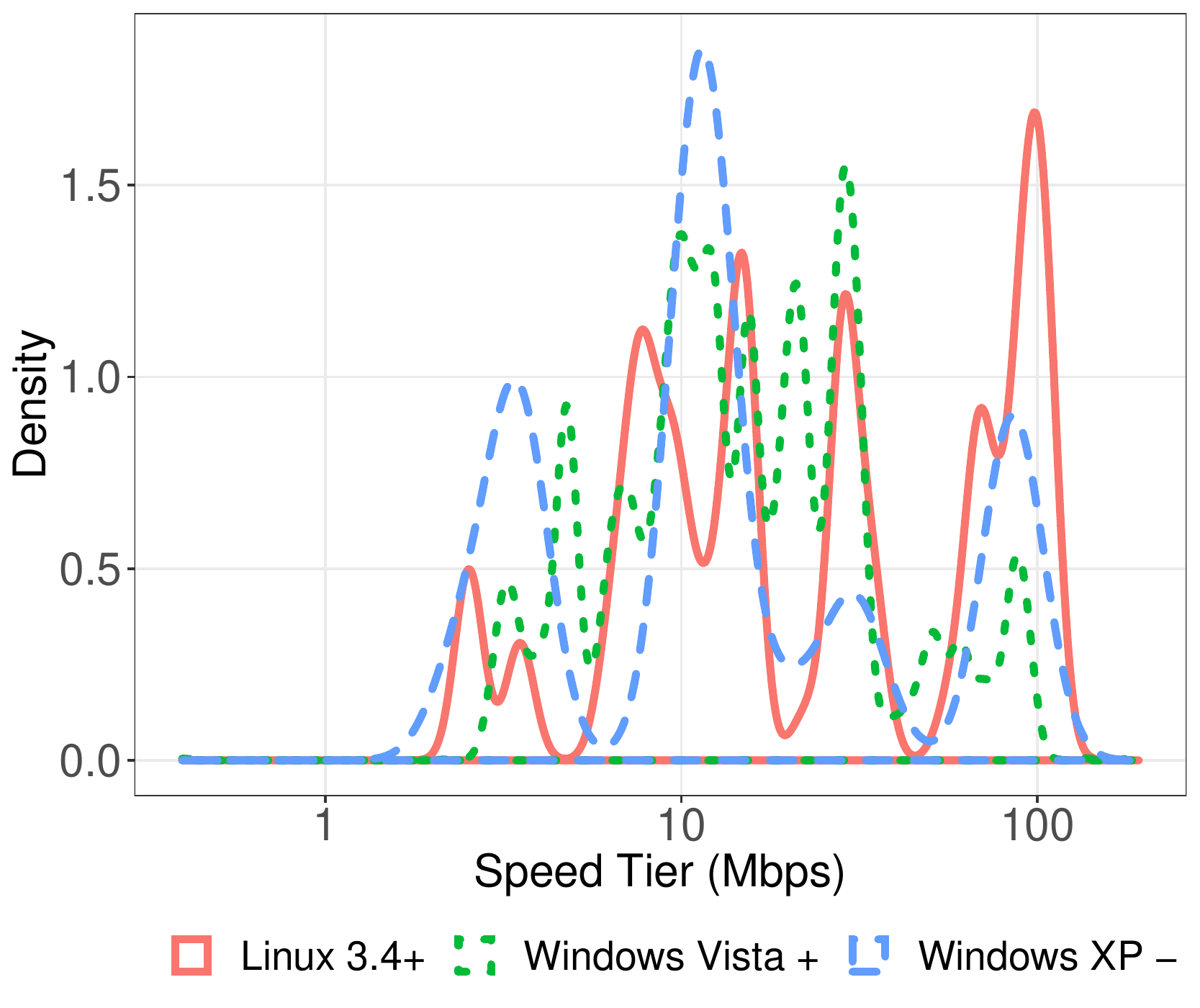}
	\vspace{-3mm}
	\caption{Comparing download speeds by OS in AU.}
	\vspace{-6mm}
	\label{fig:4ostypeAU}
\end{figure}

%
%

\subsection{Visualization Tool} \label{sec:mlab-viztool}
Since visualization is key to human comprehension and interpretation of results, we built a tool to ease the generation of plots of various performance measures (speed, RTT, congestion signals) filtered by country, ISP, or specific household, at time-scales of hours, days, and months. A {\em data-extraction} script in Python queries the M-Lab NDT store to extract data, and an R script filters fields of interest, and annotates it with extra attributes (such as speed-tier and local time) into country-specific local files. A set of {\em analytics} scripts in R then perform the various algorithmic operations, ranging from simple aggregation and normalization to more complex causal inference models discussed later. A JavaScript {\em front-end} provides user interaction to input plotting options and display the resulting graphs.

\begin{figure*}[t!]
	\begin{center}
		\mbox{
			\subfigure[AU.]{ 
				{\includegraphics[width=0.475\textwidth]{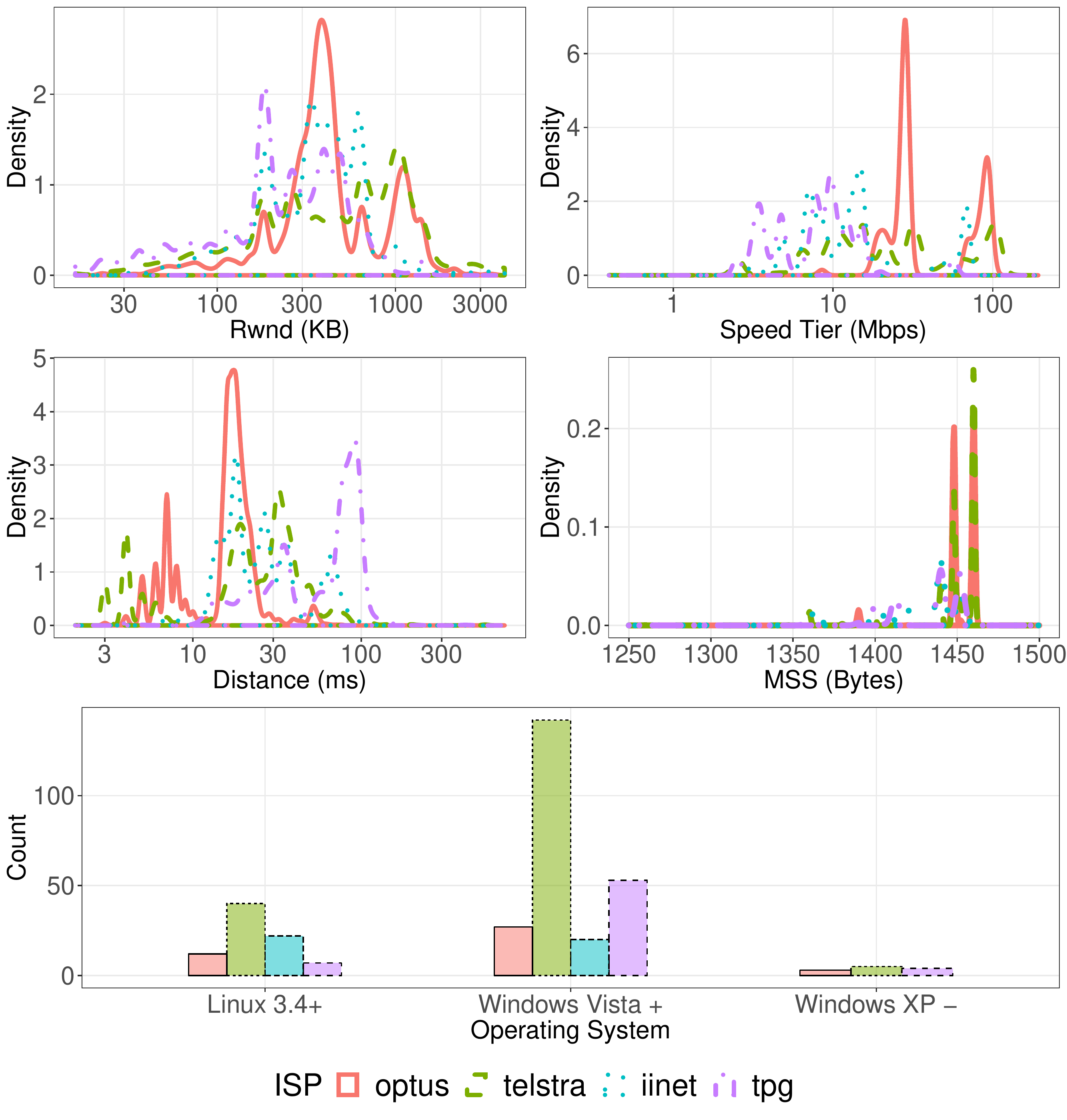}}\quad
				\label{fig:4attribdensityAU}
			}
			\hspace{-2mm}
			\subfigure[US.]{ 
				{\includegraphics[width=0.475\textwidth]{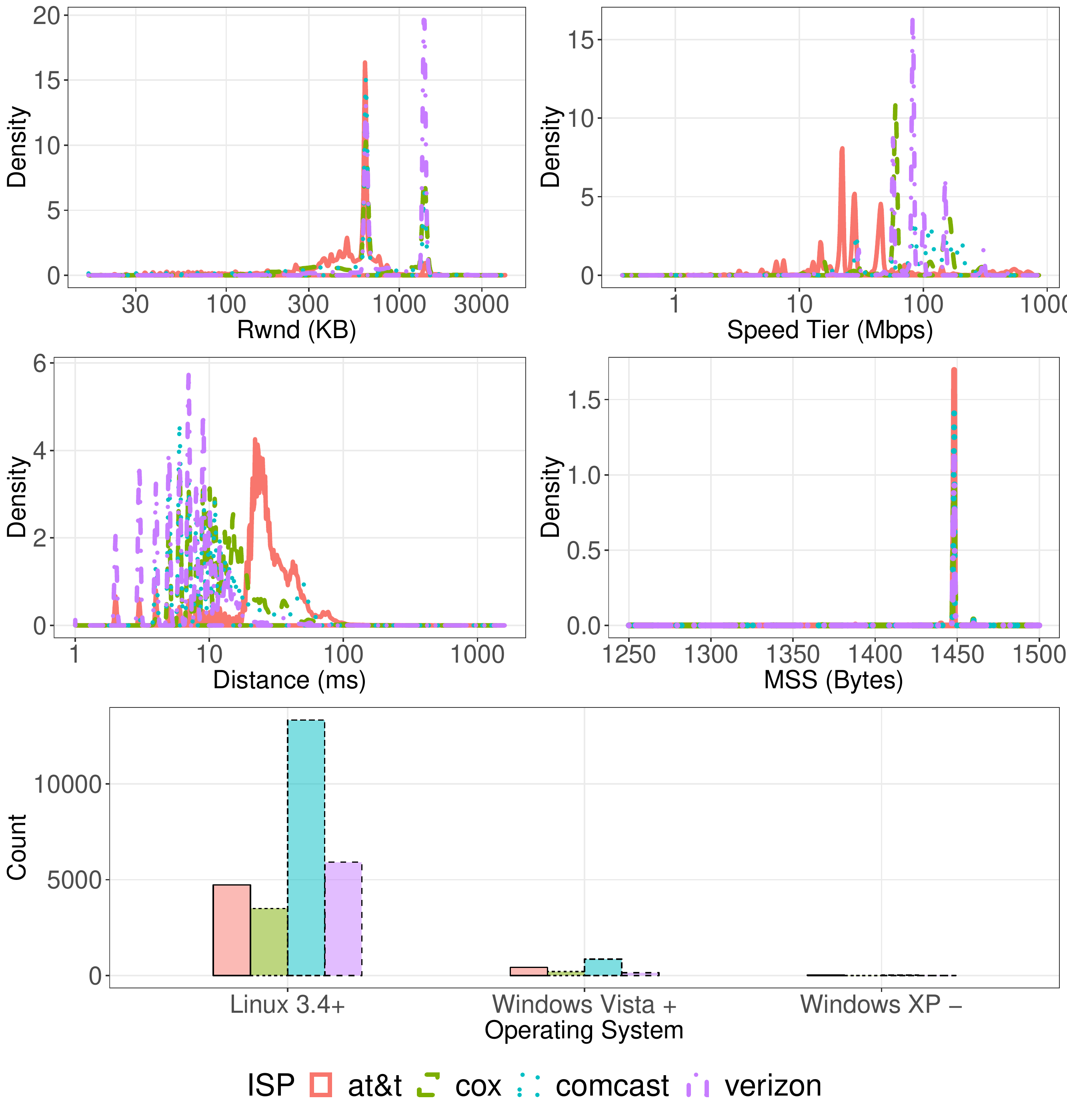}}\quad
				\label{fig:4attribdensityUS}
			}
		}
		\vspace{-4mm}
		\caption{Distribution of various attributes for: (a) AU, and (b) US.}
		\vspace{-4mm}
		\label{fig:4attribdensity}
	\end{center}
\end{figure*}

Our UI is openly accessible at {\fontsize{10}{48}\usefont{OT1}{lmtt}{b}{n}\noindent https://mlab-vis.sdn.unsw.edu.au/} and we encourage the reader to try the various plot options, such as: (a) aggregated plots that show monthly/hourly ISP (raw or normalized) speeds, (b) scatter plots that show download speeds by time-of-day or day-of-month, (c) distribution plot of speed-tier for a specific ISP, (d) correlation plot showing how the download speed relates to the various attributes, and (e) household plots that show speed, RTT, etc., specific to a client IP address. Though we will see several plots generated by our tool throughout this paper, here we would like to illustrate some visual insights from a facet plot, in this case speed measurements taken at different times of day. Fig.~\ref{fig:3facetplotUSdaily} shows a panel of 31 plots, each of which depicts all speed-tests done during that day of the month, over the month of December 2016 for Comcast. The top-10 contributing households are each given their unique color, with their IP addresses (and in parenthesis the number of test points contributed for that month) shown in the legend. One can immediately see the temporal skew in testing patterns: 
the dark green household (IP address 67.180.193.135 with 1281 tests) does its speed testing exclusively on three days (22-Dec, 29-Dec, and 30-Dec),
the light green coded household (IP address 24.147.127.89 with 672 tests) is concentrated on 18th December, while other households such as red (IP address 50.248.236.185 with  1809 tests) and purple (IP address 76.114.35.144 with 1055 tests) are spread across every day of the month.
The plot also gives a visual representation to the variability in number of tests conducted across houses, as well as the variability in speed experienced by the same household. It is often useful to corroborate the numerical results presented in this paper with their visual depictions made possible by our tool.

\begin{figure*}[t!]
	\begin{center}
		\mbox{
			\subfigure[AU.]{ 
				{\includegraphics[width=0.45\textwidth,height=0.27\textwidth]{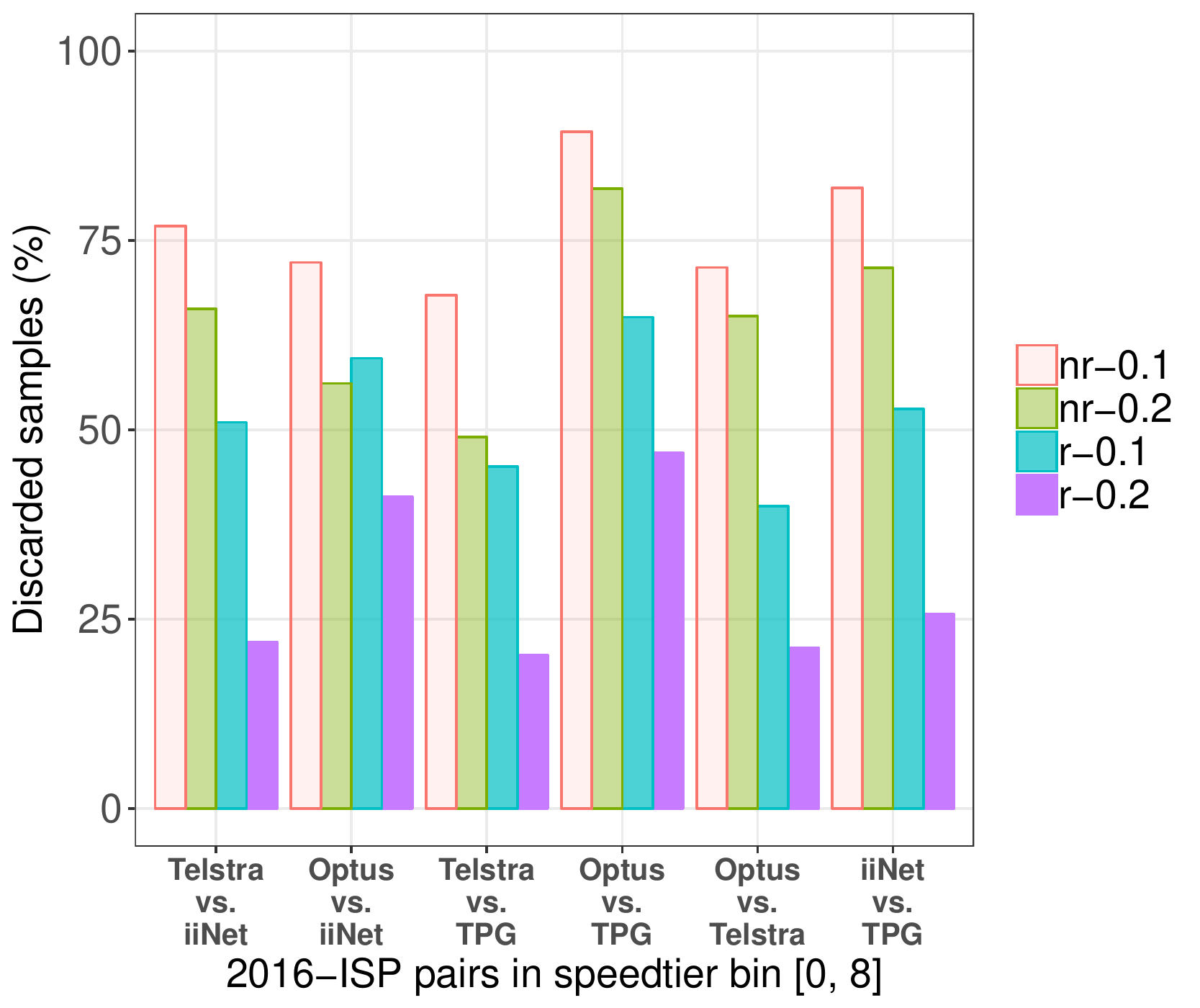}}\quad
				\label{fig:au-discard}
			}
			\hspace{-6mm}
			\subfigure[US.]{ 
				{\includegraphics[width=0.45\textwidth,height=0.27\textwidth]{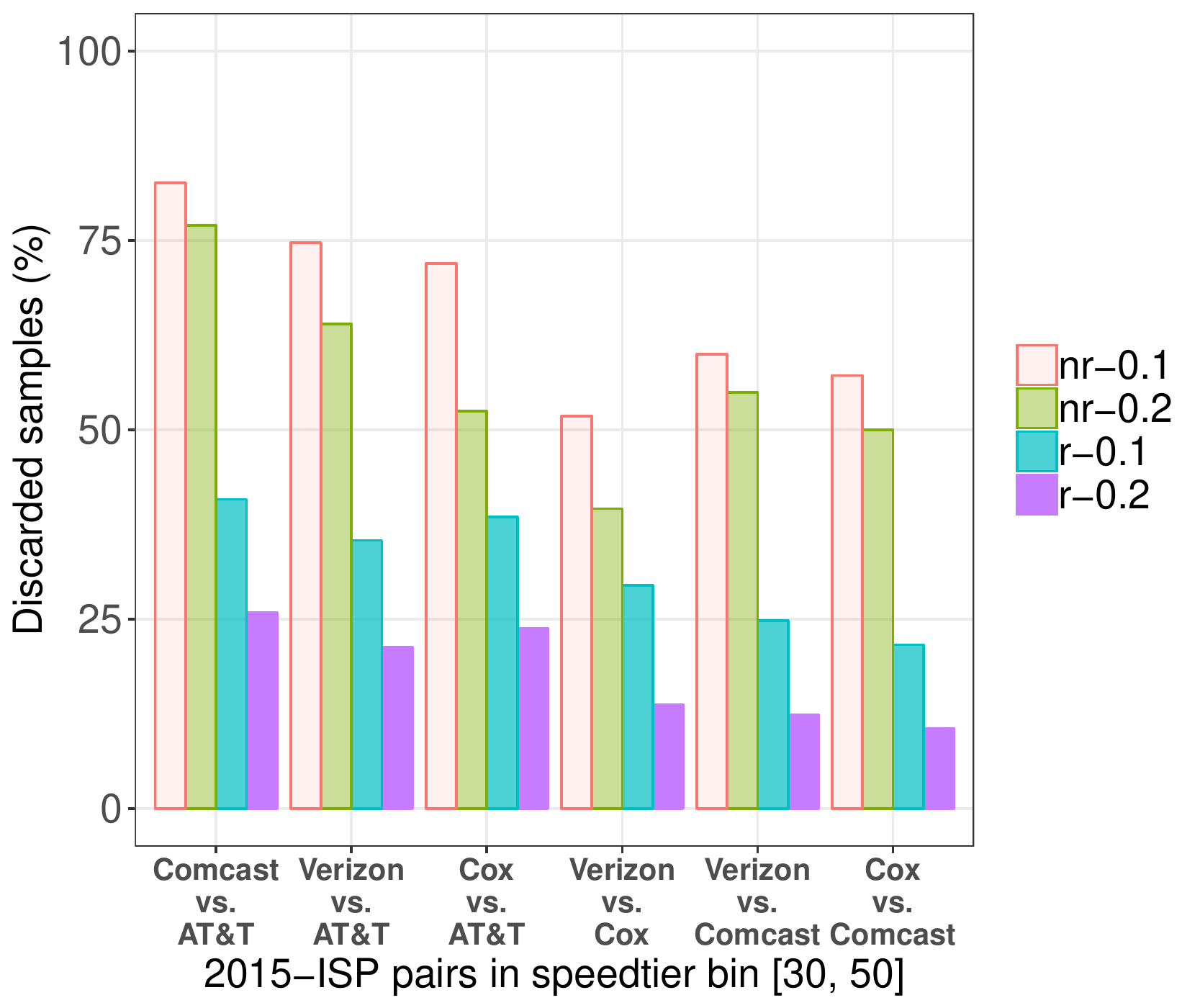}}\quad
				\label{fig:us-discard}
			}
		}
		\vspace{-3mm}
		\caption{Impact of matching with/without replacement and caliper for: (a) AU and (b) US.}
		\vspace{-7mm}
		\label{fig:discard}
	\end{center}
\end{figure*} 

\begin{figure}[t!]
	\centering
	\centering{%
		{\includegraphics[width=0.45\textwidth,height=0.75\textheight]{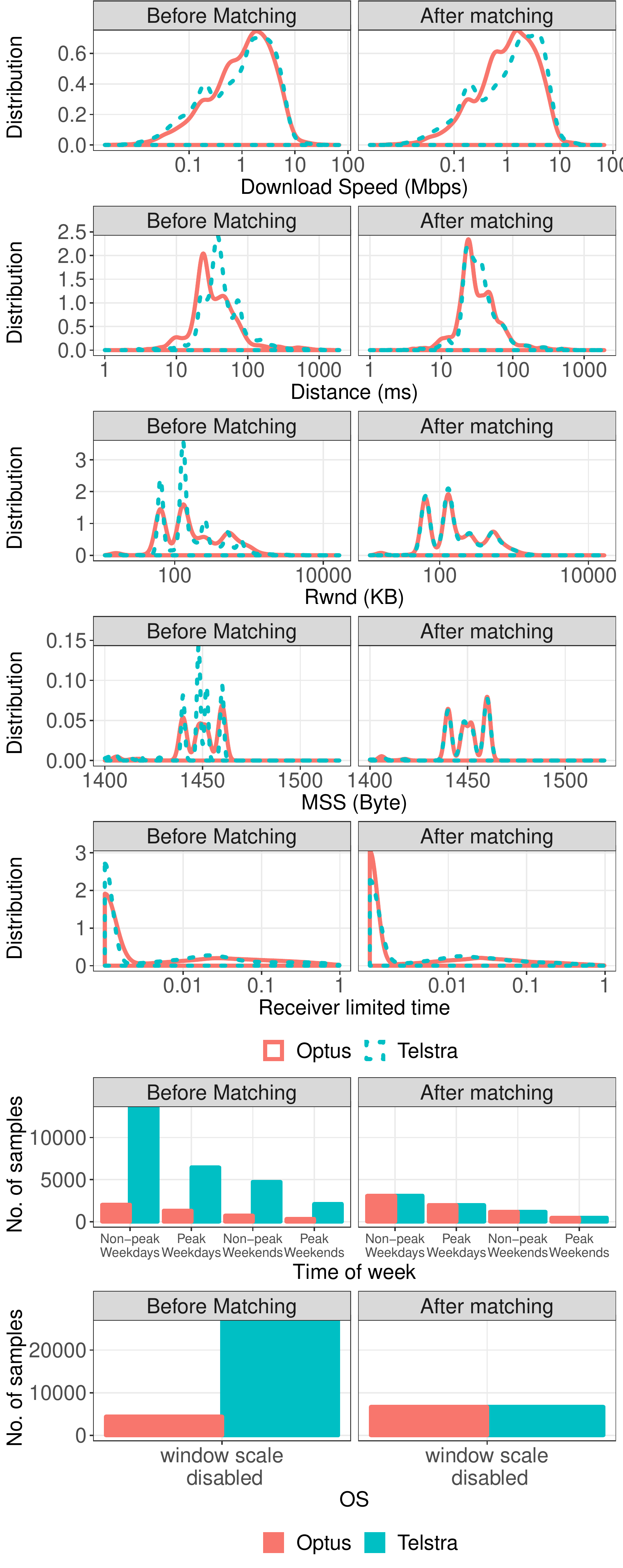}}\quad
	}
	\vspace{-5mm}	
	\caption{Covariate distribution before and after matching with replacement Optus vs. Telstra (in AU) measurements with caliper 0.2.}
	\label{fig:au-covariate}
\end{figure}

\section{Test Data Attribute Distributions and Biases} \label{sec:data-attribs}

In this section we study how test condition attributes (affecting speed) can vary in measurements across different ISPs, and how this can bias the comparison results. We begin by feeding the measurement test results, along with the associated test conditions attributes, to the {\it Random Forest} method in R to compute the ``importance rank'' (other machine learning methods for variable selection, such as Bayesian Additive Regression Trees \cite{bartvar}, provided similar results). Fig.~\ref{fig:4vimp} shows that the access speed-tier, host buffers, and distance attributes have the highest impact on measured download speed, across all the countries studied. This by itself is not very surprising, since these factors directly determine TCP dynamics and hence measured speeds. What is surprising to observe is that the ISP per-se has a relatively lower weight, typically ranking fourth or fifth in importance. Of course, in countries such as the US, each ISP often runs their own physical broadband infrastructure, and therefore wields a much larger influence via the access speed-tier attribute, whereas in countries such as AU and UK the ISPs typically share (nationalized) broadband infrastructure, and hence do not dictate the access speed-tier attribute. In either case, the ISP attribute is found to be no more significant than the operating system on the client running the test.

\begin{figure}[t!]
	\centering
	\centering{%
		{\includegraphics[width=0.45\textwidth,height=0.75\textheight]{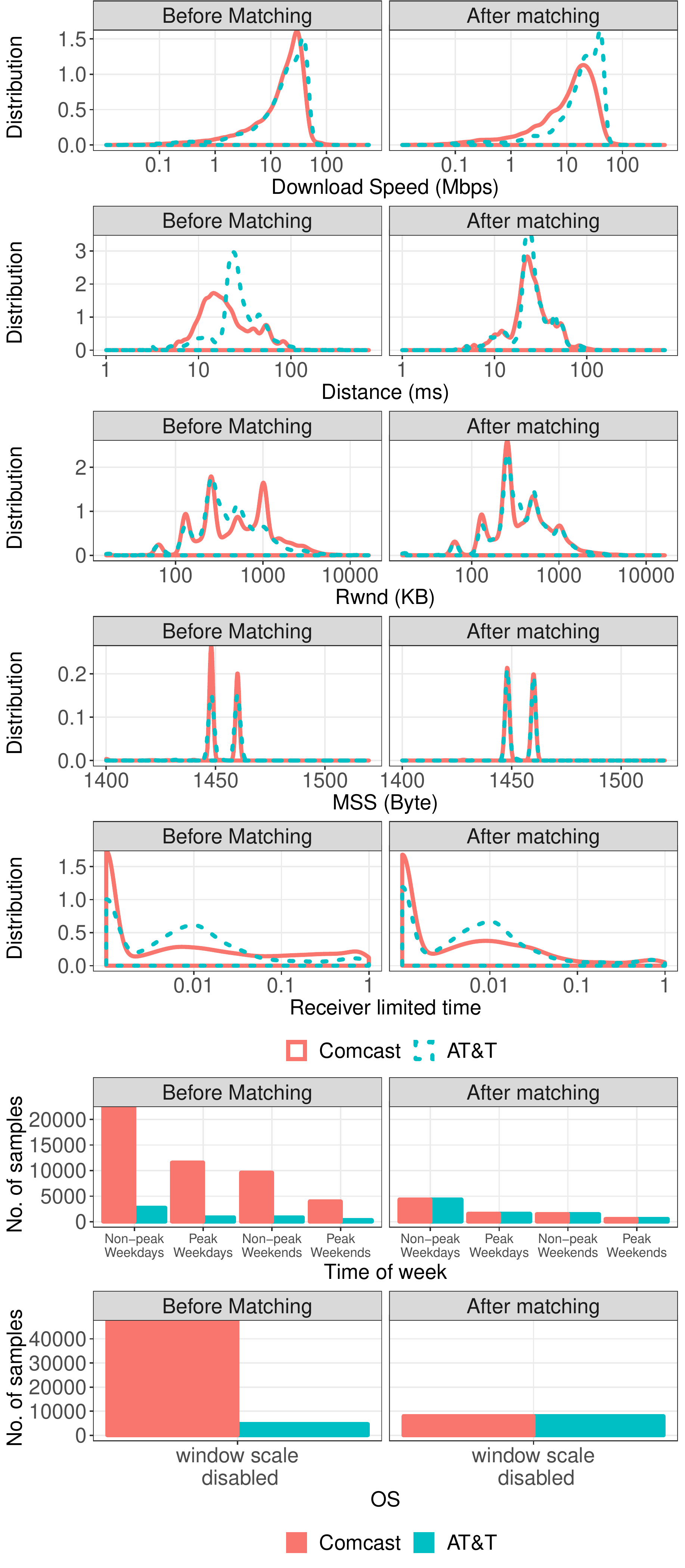}}\quad
	}
	\vspace{-5mm}
	\caption{Covariate distribution before and after matching with replacement Comcast vs. AT\&T (in US) measurements with caliper 0.2.}
	\label{fig:us-covariate}
\end{figure}

Even with a shared/nationalized broadband infrastructure, some ISPs may be serving customers with lower access speed-tiers, which can drag their averages down. In fact, in AU, Telstra claims that it serves more rural/regional customers than other ISPs such as Optus, which is used as a reason why it ranks lower on the Netflix ISP speed index (and in our monthly median plot shown in Fig.~\ref{fig:3monthlymedianAUhouse}). To check the plausibility of this claim, we use our tool to plot in Fig.~\ref{fig:4speedtierAU} the distribution of speed-tiers of households served by Tesltra and Optus in 2016. The disparity is evident -- Optus has only 10\% of subscribers at speed-tier below 8 Mbps and 63\% above 20 Mbps, while Telstra has 21\% of subscribers below 8 Mbps and only 46\% above 20 Mbps. Since much of the access infrastructure in Australia is open and can be shared by all ISPs, the disparity in access speed-tier is attributable to different proportions of metropolitan versus regional customers served by the two ISPs. Another illustration of the impact of an attribute, in this case the client OS, is shown visually in Fig.~\ref{fig:4ostypeAU}. This plot is from M-Lab test data from the AU for 2016, and shows the distribution density of measured download speed (x-axis on log scale) separated by the OS type. The bias is again evident here -- clients using flavors of Linux version 3.4 or higher (solid red curve) are clustered in the range of 10-100 Mbps, while Windows clients running XP or older (dashed blue curve) are concentrated in 5-20 Mbps -- this could be attributed to the lack of TCP auto-tuning in older versions of the Windows OS \cite{autotuning}.

For completeness, we show the distributions of all the major attributes (host TCP window size, speed-tier, distance, MSS, and OS) in Fig.~\ref{fig:4attribdensity} for AU and US (UK has similar characteristics to AU, and is omitted here for space reasons). It can be seen that in almost every attribute, the ISPs differ: for example in AU, TPG subscribers are more skewed towards lower \textit{Rwnd} (TCP window), larger client-server distance, and more Windows XP (or older) OS, compared to subscribers of other ISPs, while in the US AT\&T subscribers have larger distance and more widely spread \textit{Rwnd} than others. As discussed earlier, such differences in attributes (that are reflective of test conditions) can bias the test outcomes in multiple ways; the next section develops a method to eliminate this bias and undertake a fair comparison of ISPs.

\section{Debiasing using Multi-Variate Matching} \label{sec:matching}
We use the causal inference technique called multi-variate matching, as briefly introduced in \S\ref{sec:prior-work-causal-inference}, to balance the covariate distributions to reduce bias. For attributes that take continuous values (speed-tier, host-buffers, client-server distance, MSS), we use the {\em Mahalanobis distance} \cite{Matching11} to compute the ``closeness'' between measurement samples pertaining to ISP-1 (the ``treatment group'') and ISP-2 (the ``control group''). Samples that are within a ``caliper'' (specified in units of the standard deviation of each attribute) distance from each other are deemed to be matched, and constitute the ``common support'' between the two groups, while all other samples are dropped. A larger caliper allows more samples to be matched for greater common support, reducing error (variance) in the comparison of the average treatment effect (ATE), whereas a smaller caliper makes the matching more exact for improved unit homogeneity, yielding lower bias \cite{CausalPotential}. The caliper therefore has to be tuned to achieve the desired balance between error and bias.   

Another key factor is to check whether matching should be done ``with replacement'' (\ie \textit{r}) or ``without replacement'' (\ie \textit{nr}). Matching with replacement can often decrease bias because controls that look similar to many treated individuals can be used multiple times \cite{MatchingReview}, but this makes the inference more complex since matched controls become dependent. We employ both of these methods (\ie \textit{r} and \textit{nr}). Also for caliper, we use two values $0.1$ and $0.2$ to evaluate the impact of both tight and relax matching respectively. We show our results in Fig.~\ref{fig:discard} for six pairs of ISPs within a specific speed-tier in AU and US. 

It is seen that when matching without replacement is employed, for both AU and US, a fairly large fraction of samples (\ie typically between $50$\% to $75$\%) fall out of the common support, and thus are discarded -- except for ``Verizon vs. Cox'' pair in US with the caliper value $0.1$. Instead, matching with replacement discards less fraction of samples (or keeps a larger portion of samples in common support area). Focusing on AU in Fig.~\ref{fig:au-discard}, the discard rate is less than $25$\% for four pairs of ISPs when matching with replacement is employed and the caliper value is set to $0.2$ (purple bars) -- two pairs experience a relatively higher discard rate about $45$\%.  For US, on the other hand, same settings give a fairly consistent discard rate of less than $25$\% across four pairs of ISPs, as shown by purple bars in Fig~\ref{fig:us-discard}. As a results, we use matching with replacement and caliper $0.2$ having sufficient number of samples in the common support to keep error low.   


\begin{table*}[!t]
	\centering
	\caption{ISP pairs with common support in AU and US.}
	\label{tab:ISPpairs}
	\begin{adjustbox}{max width=0.995\textwidth}
		\renewcommand{\arraystretch}{1.3}      	
		\begin{tabular}{|l|c|c|c|c|c|c|c|c|c|c|c|c|c|c|c|c|c|c|c|c|c|}
			\hline 
			\textbf{ISP pair ID} & \textbf{1} & \textbf{2} & \textbf{3} & \textbf{4} & \textbf{5} & \textbf{6} & \textbf{7} & \textbf{8} & \textbf{9} & \textbf{10} & \textbf{11} & \textbf{12} & \textbf{13} & \textbf{14} & \textbf{15} & \textbf{16} & \textbf{17} & \textbf{18} & \textbf{19} & \textbf{20} & \textbf{21}\tabularnewline
			\hline 
			\textbf{ISP1 vs ISP2} &  \rotatebox{90}{Comcast vs. AT\&T} &  \rotatebox{90}{Comcast vs. AT\&T} &  \rotatebox{90}{Virgin vs. BT} &  \rotatebox{90}{Comcast vs. Verizon} &  \rotatebox{90}{Comcast vs. Verizon} &  \rotatebox{90}{Comcast vs. Cox} &  \rotatebox{90}{Comcast vs. AT\&T} &  \rotatebox{90}{Virgin vs. BT} &  \rotatebox{90}{Comcast vs. AT\&T} &  \rotatebox{90}{Virgin vs. BT} &  \rotatebox{90}{Virgin vs. BT} &  \rotatebox{90}{Comcast vs. Cox} &  \rotatebox{90}{Comcast vs. Cox} &  \rotatebox{90}{Comcast vs. Cox} &  \rotatebox{90}{Virgin vs. Sky} &  \rotatebox{90}{Comcast vs. AT\&T} &  \rotatebox{90}{Virgin vs. Sky} &  \rotatebox{90}{Virgin vs. BT} & \rotatebox{90}{ Virgin vs. Sky} &  \rotatebox{90}{Virgin vs. Sky} &  \rotatebox{90}{Virgin vs. BT}\tabularnewline
			\hline 
			\textbf{Speed-tier} &  {[}30, 50{]} &  {[}30, 50{]} &  {[}50, 75{]} &  {[}75, 100{]} &  {[}20, 25{]} &  {[}50, 75{]} &  {[}75, 100{]} &  {[}25, 30{]} &  {[}20, 25{]} &  {[}30, 50{]} &  {[}30, 50{]} &  {[}30, 50{]} &  {[}20, 25{]} &  {[}12, 20{]} &  {[}25, 30{]} &  {[}50, 75{]} &  {[}30, 50{]} &  {[}12, 20{]} &  {[}20, 25{]} &  {[}25, 30{]} &  {[}20, 25{]}\tabularnewline
			\hline 
			\textbf{Year} & 2016 & 2015 & 2015 & 2015 & 2015 & 2015 & 2016 & 2016 & 2016 & 2015 & 2016 & 2016 & 2016 & 2016 & 2016 & 2016 & 2016 & 2015 & 2015 & 2015 & 2015\tabularnewline
			\hline\hline 
			
			\textbf{ISP pair ID} & \textbf{22} & \textbf{23} & \textbf{24} & \textbf{25} & \textbf{26} & \textbf{27} & \textbf{28} & \textbf{29} & \textbf{30} & \textbf{31} & \textbf{32} & \textbf{33} & \textbf{34} & \textbf{35} & \textbf{36} & \textbf{37} & \textbf{38} & \textbf{39} & \textbf{40} & \textbf{41} & \textbf{42}\tabularnewline
			\hline 
			\textbf{ISP1 vs ISP2} &  \rotatebox{90}{Virgin vs. TalkTalk} &  \rotatebox{90}{Telstra vs. Optus} &  \rotatebox{90}{Comcast vs. AT\&T} &  \rotatebox{90}{Comcast vs. AT\&T} &  \rotatebox{90}{Comcast vs. Cox} &  \rotatebox{90}{Virgin vs. Sky} &  \rotatebox{90}{Comcast vs. AT\&T} &  \rotatebox{90}{Comcast vs. Verizon} &  \rotatebox{90}{Virgin vs. Sky} &  \rotatebox{90}{Comcast vs. Verizon} &  \rotatebox{90}{Virgin vs. Sky} &\rotatebox{90}{Comcast vs. Verizon} &  \rotatebox{90}{Comcast vs. AT\&T} &  \rotatebox{90}{Comcast vs. AT\&T} &  \rotatebox{90}{Comcast vs. Verizon} &  \rotatebox{90}{TPG vs. Optus} &  \rotatebox{90}{Comcast vs. AT\&T} & \rotatebox{90}{TPG vs. Telstra} &  \rotatebox{90}{Comcast vs. Cox} &  \rotatebox{90}{Comcast vs. Verizon} &  \rotatebox{90}{Comcast vs. Cox} \tabularnewline
			\hline 
			\textbf{Speed-tier} &  {[}12, 20{]} & {[}25, 30{]}& {[}12, 20{]} &  {[}25, 30{]} &  {[}30, 50{]} &  {[}12, 20{]} &  {[}0, 8{]} &  {[}50, 57{]} &  {[}30, 50{]} &  {[}30, 50{]} &  {[}12, 20{]} &  {[}20, 25{]} &  {[}8, 12{]} &  {[}20, 25{]} &  {[}25, 30{]} &  {[}12, 20{]} &  {[}0, 8{]} &  {[}12, 20{]} &  {[}25, 30{]} &  {[}50, 57{]} &  {[}20, 25{]}\tabularnewline
			\hline 
			\textbf{Year} & 2016 & 2016 & 2016 & 2015 & 2015 & 2016 & 2016 & 2016 & 2015 & 2016 & 2015 & 2016 & 2016 & 2015 & 2015 & 2015 & 2015 & 2015 & 2015 & 2015 & 2015\tabularnewline
			\hline\hline 
			
			\textbf{ISP pair ID} & \textbf{43} & \textbf{44} & \textbf{45} & \textbf{46} & \textbf{47} & \textbf{48} & \textbf{49} & \textbf{50} & \textbf{51} & \textbf{52} & \textbf{53} & \textbf{54} & \textbf{55} & \textbf{56} & \textbf{57} & \textbf{58} & \textbf{59} & \textbf{60} & \textbf{61} & \textbf{62} & \textbf{63}\tabularnewline
			\hline 
			\textbf{ISP1 vs ISP2} &  \rotatebox{90}{Comcast vs. AT\&T} &  \rotatebox{90}{Comcast vs. Cox} &  \rotatebox{90}{Comcast vs. AT\&T} &  \rotatebox{90}{Comcast vs. Verizon} &  \rotatebox{90}{Comcast vs. Cox} &  \rotatebox{90}{Comcast vs. Verizon} &  \rotatebox{90}{Comcast vs. Verizon} &  \rotatebox{90}{Comcast vs. Cox} &  \rotatebox{90}{Comcast vs. AT\&T} &  \rotatebox{90}{Comcast vs. Verizon} &  \rotatebox{90}{Comcast vs. Verizon} &  \rotatebox{90}{Telstra vs. iiNet} &  \rotatebox{90}{TPG vs. Telstra} &  \rotatebox{90}{Telstra vs. iiNet} &  \rotatebox{90}{Telstra vs. Optus} &  \rotatebox{90}{Comcast vs. Cox} &  \rotatebox{90}{TPG vs. Optus} &  \rotatebox{90}{Telstra vs. TPG} & \rotatebox{90}{Comcast vs. Cox} &  \rotatebox{90}{Telstra vs. iiNet} &  \rotatebox{90}{Telstra vs. TPG}\tabularnewline
			\hline 
			\textbf{Speed-tier} &  {[}8, 12{]} &  {[}8, 12{]} &  {[}12, 20{]} &  {[}30, 50{]} &  {[}12, 20{]} &  {[}12, 20{]} &  {[}0, 8{]} &  {[}0, 8{]} &  {[}25, 30{]} &  {[}0, 8{]} &  {[}8, 12{]} &  {[}0, 8{]} &  {[}8, 12{]} &  {[}8, 12{]} &  {[}0, 8{]} &  {[}8, 12{]} &  {[}0, 8{]} &  {[}0, 8{]} &  {[}75, 100{]} &  {[}12, 20{]} &  {[}8, 12{]}\tabularnewline
			\hline 
			\textbf{Year} & 2015 & 2015 & 2015 & 2015 & 2015 & 2015 & 2016 & 2015 & 2016 & 2015 & 2015 & 2016 & 2015 & 2016 & 2016 & 2016 & 2015 & 2016 & 2015 & 2016 & 2016\tabularnewline
			\hline\hline

			\textbf{ISP pair ID} & \textbf{64} & \textbf{65} & \textbf{66} & \textbf{67} & \textbf{68} & \textbf{69} & \textbf{70} & \textbf{71} & \textbf{72} & \textbf{73} & \textbf{74} & \textbf{75} & \textbf{76} & \textbf{77} & \textbf{78} & \textbf{79} & \textbf{80} & \textbf{81} & \textbf{82} &  \textbf{83} & \tabularnewline
			\hline 
			\textbf{ISP1 vs ISP2} &  \rotatebox{90}{Comcast vs. Verizon} &  \rotatebox{90}{Comcast vs. Cox} &  \rotatebox{90}{Telstra vs. iiNet} &  \rotatebox{90}{Virgin vs. Sky} &  \rotatebox{90}{Telstra vs. Optus} &  \rotatebox{90}{Telstra vs. Optus} &  \rotatebox{90}{Telstra vs. Optus} &  \rotatebox{90}{Virgin vs. Sky} &  \rotatebox{90}{Telstra vs. TPG} &  \rotatebox{90}{Virgin vs. BT} &  \rotatebox{90}{TPG vs. Telstra} &  \rotatebox{90}{TPG vs. iiNet} &  \rotatebox{90}{Comcast vs. Verizon} &  \rotatebox{90}{Comcast vs. Cox} &  \rotatebox{90}{Telstra vs. iiNet} &  \rotatebox{90}{Telstra vs. Optus} &  \rotatebox{90}{Telstra vs. Optus} &  \rotatebox{90}{Comcast vs. Cox} & \rotatebox{90}{ Comcast vs. Verizon} & \rotatebox{90}{Comcast vs. Cox} &  \tabularnewline
			\hline 
			\textbf{Speed-tier} &  {[}8, 12{]} &  {[}0, 8{]} &  {[}20, 25{]} &  {[}0, 8{]} &  {[}8, 12{]} &  {[}12, 20{]} &  {[}20, 25{]} &  {[}0, 8{]} &  {[}12, 20{]} &  {[}0, 8{]} &  {[}0, 8{]} &  {[}0, 8{]} &  {[}12, 20{]} &  {[}100, 1000{]} &  {[}30, 50{]} &  {[}30, 50{]} &  {[}50, 75{]} &  {[}75, 100{]} &  {[}25, 30{]} &  {[}25, 30{]} &  \tabularnewline
			\hline 
			\textbf{Year} & 2016 & 2016 & 2016 & 2016 & 2016 & 2016 & 2016 & 2015 & 2016 & 2015 & 2015 & 2015 & 2016 & 2015 & 2016 & 2016 & 2016 & 2016 & 2016 & 2016 & \tabularnewline
			\hline								
			
		\end{tabular}		
	\end{adjustbox}
\end{table*}	

\subsection{Matching on Selected ISP Pairs}
Let us begin by looking at Optus vs Telstra in AU for speed-tier [$0$, $8$] Mbps. In Fig.~\ref{fig:au-covariate}, we show how covariate measures and download speeds change when matching with replacement of caliper $0.2$ is applied. It is seen that attributes differ substantially before and after matching -- matching discards $20.8$\% of samples and makes the covariate distributions more similar (thereby reducing bias). Note that, before matching, Telstra is disadvantaged in terms of longer distance and larger number of users with old OS (not supporting TCP auto-tuning), and hence sees a smaller download speed on average, suggested by naive aggregation (\eg Fig.~\ref{fig:3monthlymedianAUhouse}). When confounding factors (attributes) are balanced using the matching method, interestingly Telstra gives better performance than Optus on average -- a result contrary to the one seen in Fig.~\ref{fig:3monthlymedianAUhouse}.

We now compare Comcast and AT\&T in the US for speed-tier bin [$30$, $50$] Mbps. We can observe in Fig.~\ref{fig:us-covariate} that attributes are quite disparate between Comcast and AT\&T when all samples are considered (before matching). Comcast is disadvantaged against AT\&T by several factors including: a) a higher distance between clients and server; b) a larger number of measurements from old OS (\ie window scale disabled); and c) a larger number of measurements in the state with high value (\ie $0.1$ to $1$) of Receiver-Limited-Time. Therefore, a naive average difference suggests that Comcast is about $8$ Mpbs slower than AT\&T on average. But matching balances the attributes (confounding factors) as shown by plots on the right in Fig.~\ref{fig:us-covariate} and thus reduces the speed difference to $3$-$4$ Mbps.

%

\begin{figure}[!t]
	\centering
	\includegraphics[width=0.5\textwidth]{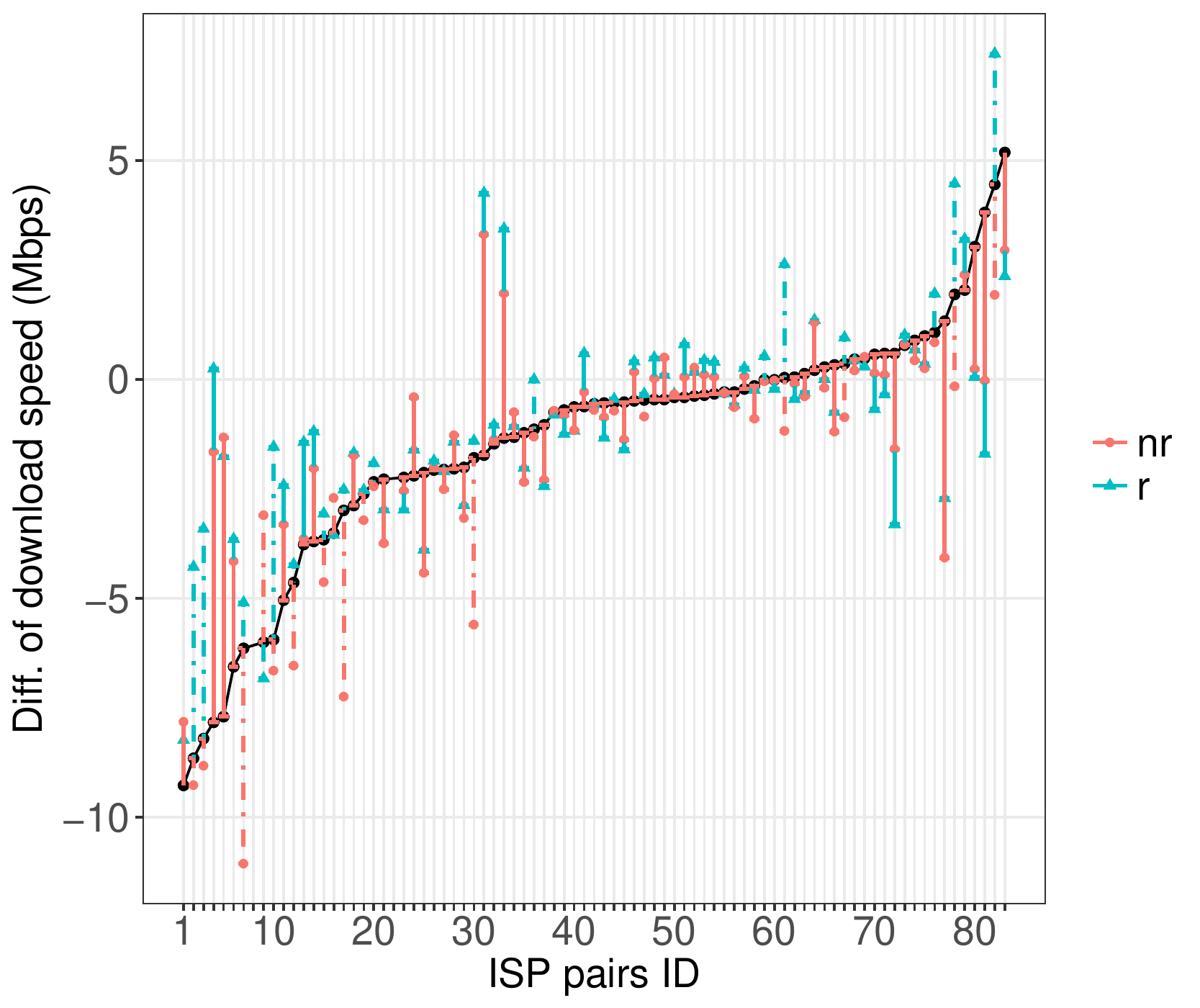}
	\vspace{-6mm}
	\caption{Comparing ISP speeds for 83 pairs in AU and US.}
	\vspace{-4mm}
	\label{fig:ISPpairs}
\end{figure}

\begin{figure*}[!t]
	\begin{center}
		\mbox{
			\subfigure[Negative correlation (Cox, 458 tests from {\fontsize{8}{48}\usefont{OT1}{lmtt}{b}{n}\noindent 98.174.39.22}) -- high speed during un-congested period, and low speed during fairly congested period.]{
				{\includegraphics[width=0.45\textwidth,height=0.23\textheight]{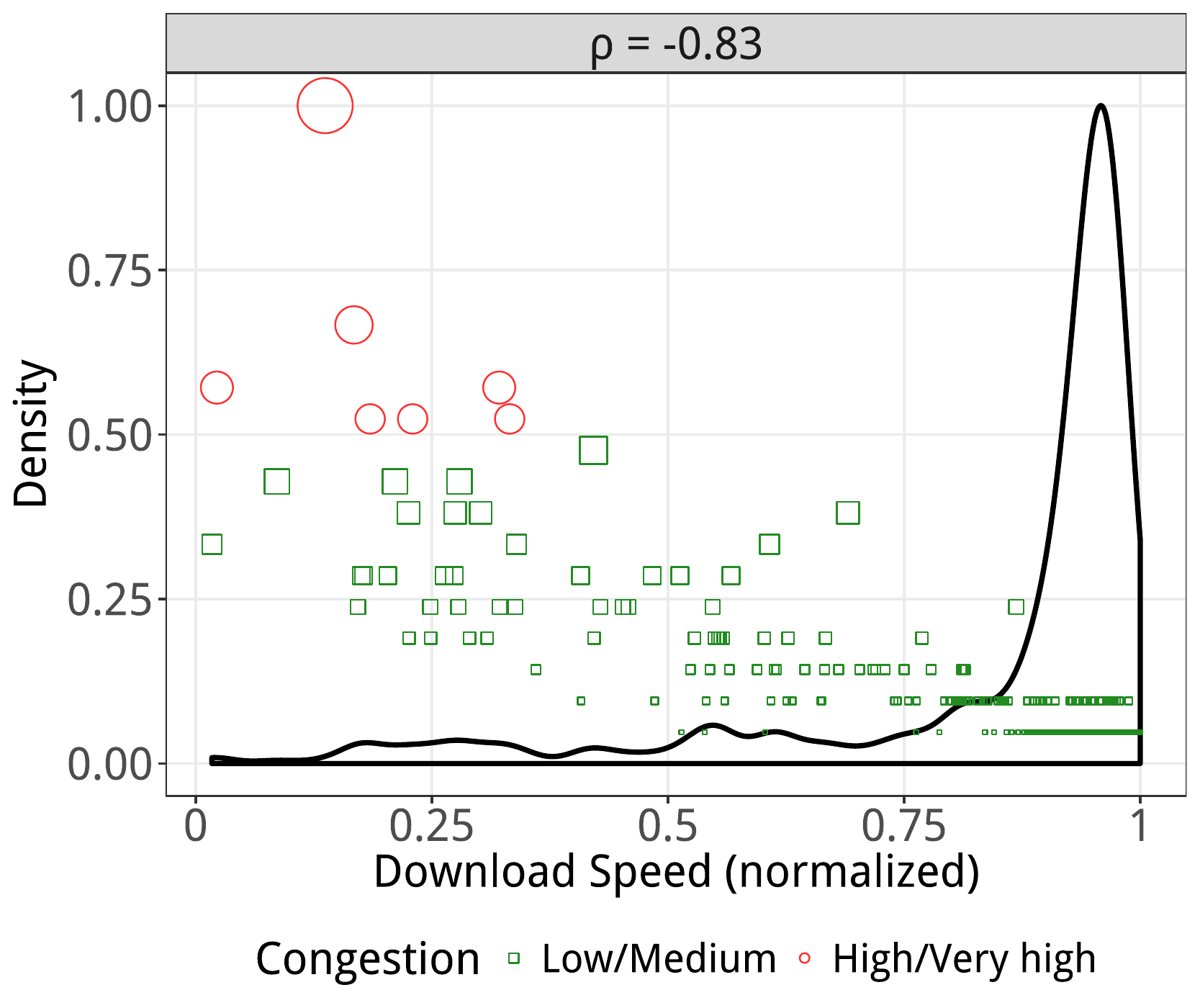}}\quad
				\label{fig:NegCorrAgg}
			}
			\hspace{0mm}
			\subfigure[Positive correlation (City of Thomasville Utilities), 896 tests from {\fontsize{8}{48}\usefont{OT1}{lmtt}{b}{n}\noindent 64.39.155.194}) -- high speed even during highly congested period, and low speed even during uncongested period.]{
				{\includegraphics[width=0.45\textwidth,height=0.23\textheight]{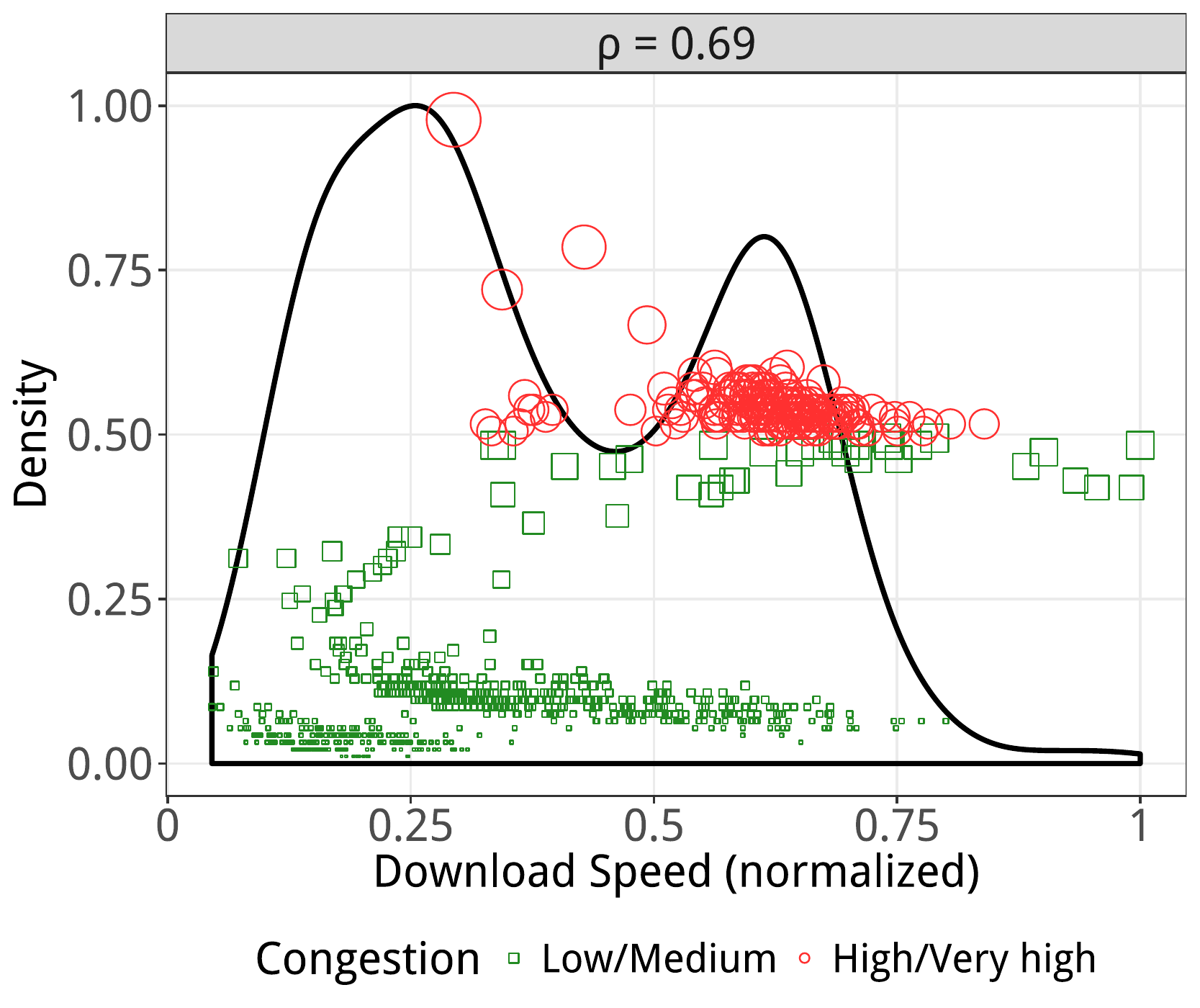}}\quad
				\label{fig:PosCorrAgg}
			}
		}
		\vspace{-3mm}
		\caption{Two samples of correlation between download-speed and congestion-count.}
		\vspace{-7mm}
		\label{fig:Corr}
	\end{center}
\end{figure*}

\subsection{Matching on Larger Set of ISP Pairs and Various Speed-Tiers}
We now extend our evaluation to a larger set of ISP pairs with diversity of speed-tiers for dataset from two years 2015 and 2016. Table~\ref{tab:ISPpairs} shows 83 ISP pairs (in AU and US) along with their corresponding speed-tier (in Mbps) and the year of data. ISP pairs are sorted in ascending order of their speed difference (computed from naive aggregation) -- each pair is assigned a unique ID (order) for ease of reference in Table~\ref{tab:ISPpairs}. 

We show the results of matching with caliper $0.2$ in Fig.~\ref{fig:ISPpairs}. The speed difference of each ISP pair resulted from naive arithmetic means, matching without replacement (\textit{nr}), and matching with replacement (\textit{r})  are shown by black dots, red segments, and green segments respectively. Matching estimated differences are represented by error bars of 95\% confidence interval.

We make the following observations: (a) when raw difference is close to zero for pair IDs between 40 to 70, the estimated differences after matching are also close to zero -- matching method does not change the inference, (b) when raw difference has a large negative value for pair IDs between 1 to 20 (or a large positive value for pair IDs between 75 to 83), matching estimates the difference much lower (\ie 0 to -3 Mbps) than what naive arithmetic means indicate (\ie -3 to -8 Mbps), and (c) in some cases, the raw difference is positive but the estimated value becomes negative with statistical significance -- this indicates that a simple use of raw average for ranking ISPs could be misleading, for example Optus vs. Telstra in AU (as discussed earlier).


\section{Refining Speed-Tier Estimation} \label{sec:refinedMatching}
We developed a causal inference model for a fair comparison of ISP performance in the previous section. In this section we refine our model by undertaking a more detailed analysis to estimate household speed-tier (also referred to as ``broadband access capacity'') from the MLab data, allowing us to further improve the fidelity of broadband speed comparisons.


\begin{figure*}[t!]
	\begin{center}
		\mbox{
			\subfigure[AU.]{
				{\includegraphics[width=0.45\textwidth,height=0.23\textheight]{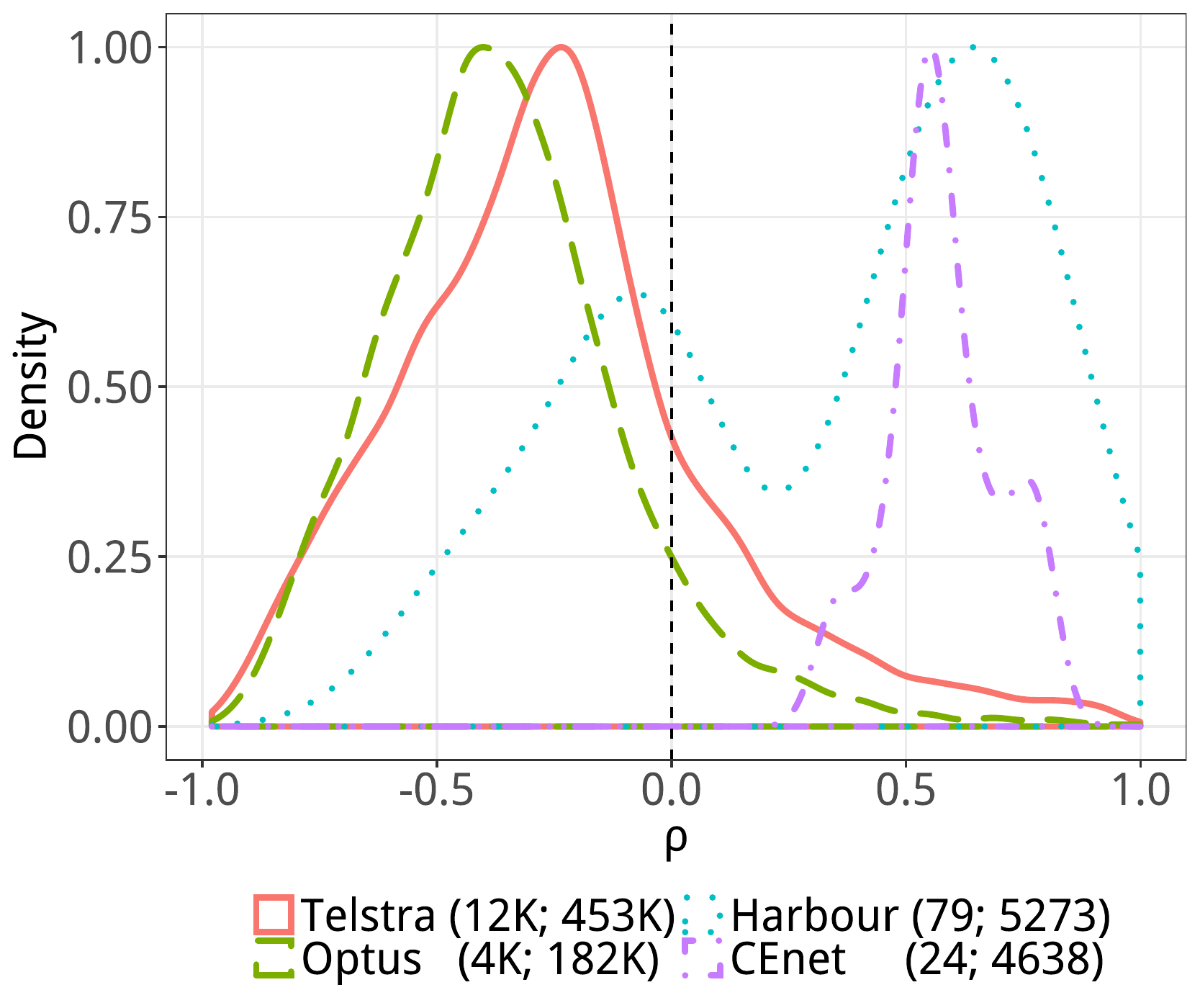}}\quad
				\label{fig:rhoAU}
			}
			\hspace{-2mm}
			\subfigure[US.]{
				{\includegraphics[width=0.45\textwidth,height=0.23\textheight]{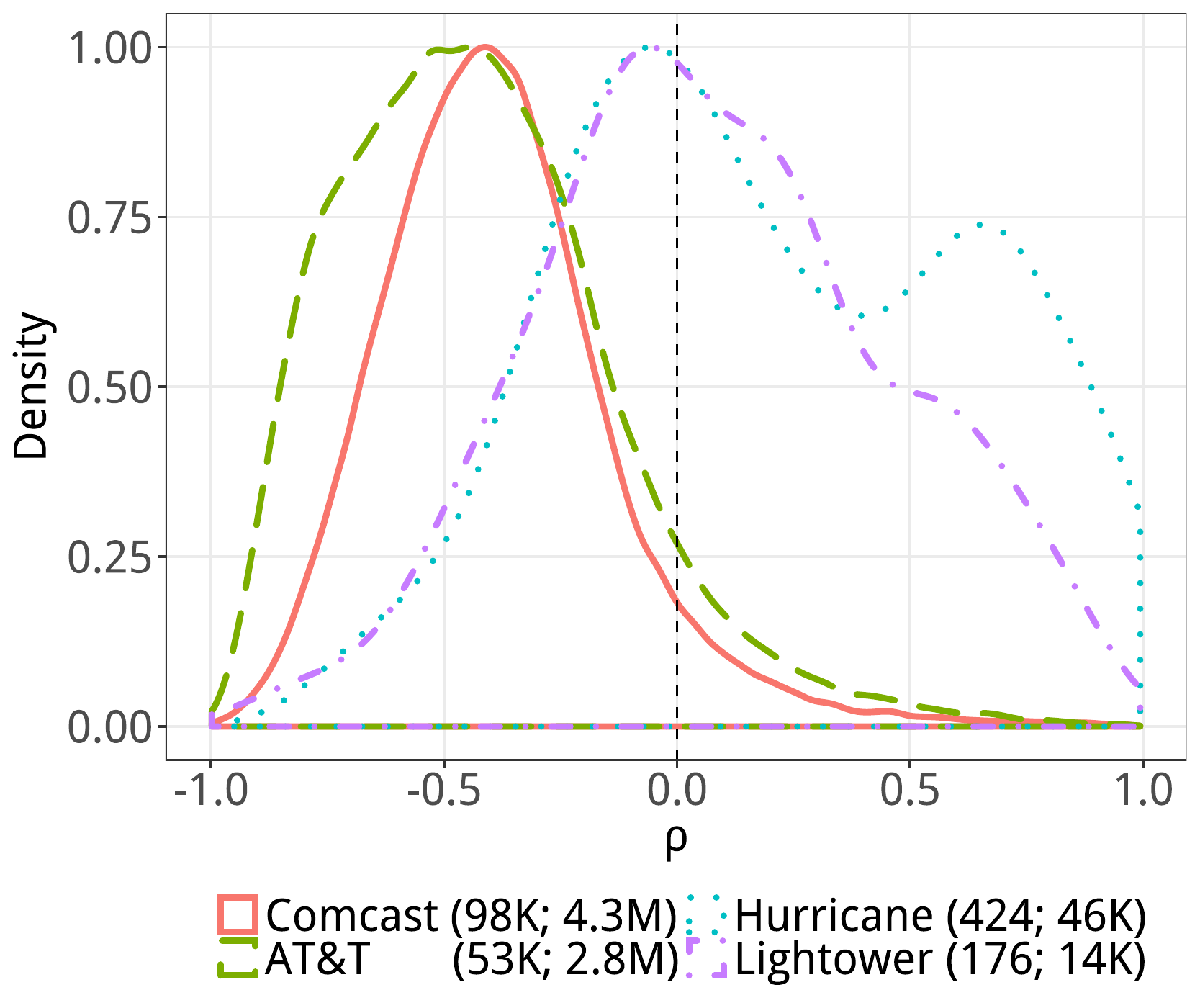}}\quad
				\label{fig:rhoUS}
			}
		}
		\vspace{-3mm}
		\caption{ Negative/Positive correlation across large/small ISPs in (a) AU, and (b) US.}
		\vspace{-5mm}
		\label{fig:rhoAUUS}
	\end{center}
\end{figure*}

\begin{figure*}[t!]
	\begin{center}
		\mbox{
			\subfigure[Negative correlation (Cox, 458 tests from {\fontsize{8}{48}\usefont{OT1}{lmtt}{b}{n}\noindent 98.174.39.22}).]{
				{\includegraphics[width=0.45\textwidth,height=0.23\textheight]{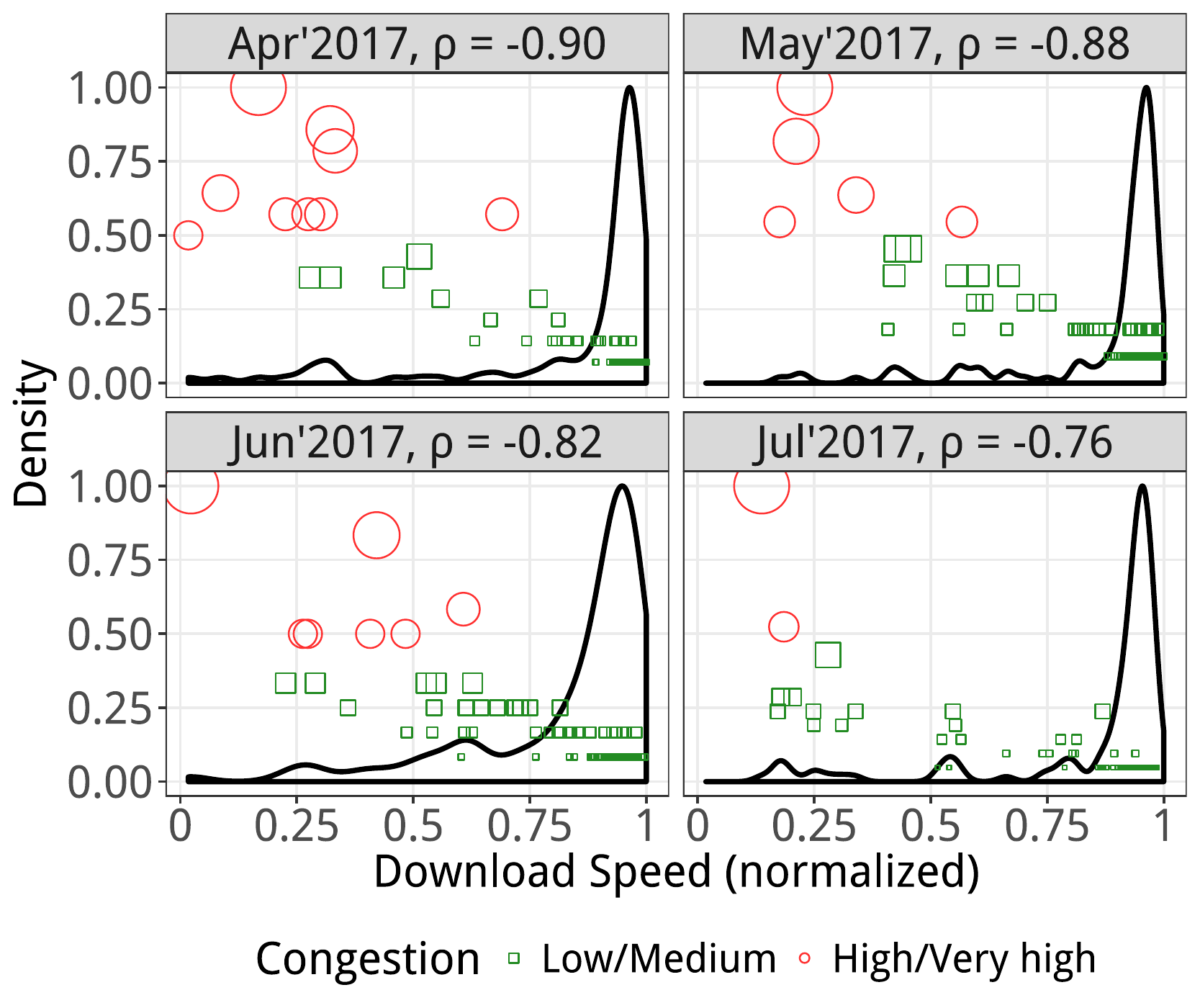}}\quad
				\label{fig:NegCorrMnt}
			}
			\hspace{-2mm}
			\subfigure[Positive correlation (City, 896 tests from {\fontsize{8}{48}\usefont{OT1}{lmtt}{b}{n}\noindent 64.39.155.194}).]{
				{\includegraphics[width=0.45\textwidth,height=0.23\textheight]{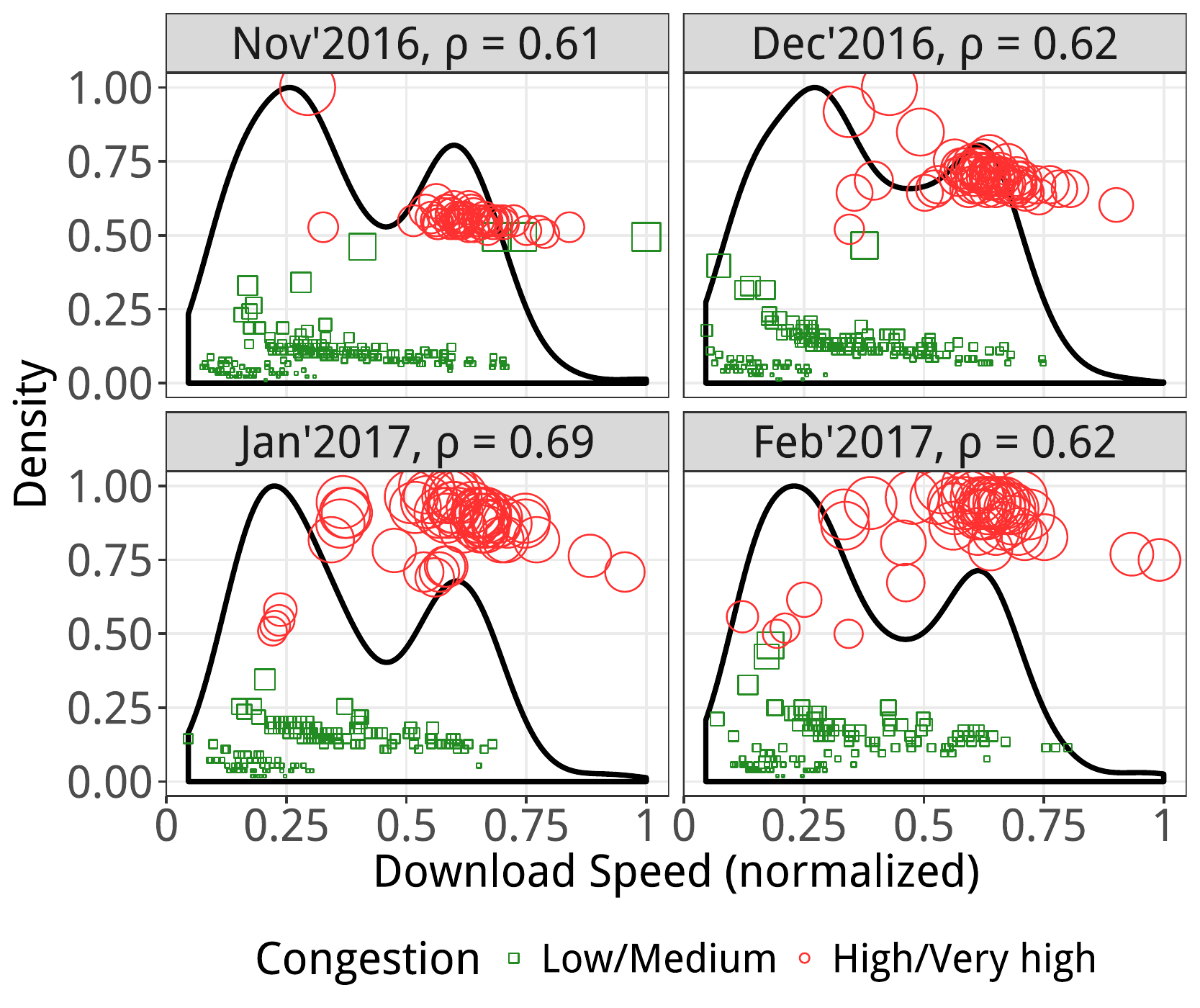}}\quad
				\label{fig:PosCorrMnt}
			}
		}
		\vspace{-3mm}
		\caption{Consistency of correlation between download-speed and congestion-count across four months: (a) negative, and (b) positive correlation.}
		\vspace{-5mm}
		\label{fig:CorrMnt}
	\end{center}
\end{figure*}

\subsection{Isolating Households}\label{sec:isolation}

M-Lab data-points are indexed by IP address of home gateways. ISPs allocate IP addresses based on their resource pool, subscriber base, or business policy. In some cases, an ISP (often a large one) may have a fairly large pool of public IP addresses and can assign every subscriber a unique public IP, but one-to-one address lease may change dynamically over time. In other cases, the ISP (often a small one) will instead assign a public IP address to a group of subscribers, and then employ NAT to multiplex their traffic. Consequently, it becomes challenging to extract the broadband capacity from M-Lab data, as an IP address does not necessarily represent a single household. Thus, we need a method to isolate data-points corresponding to single households.

The congestion signal of each NDT data-point indicates how the TCP congestion window (\textit{cwnd}) is being affected by congestion and is incremented by any form of congestion notifications including Fast Retransmit, ECN, and timeouts. Theoretically, a large value of congestion signal (congestion count) should correspond to a low TCP throughput (download speed), and vice versa. 

We denote the \textbf{Pearson's correlation coefficient} between the measured download speed and recorded congestion count by $\rho$. This parameter is computed across all tests corresponding to a given client IP address.

We expect $\rho$ to be negative for any given household, as higher broadband speed should correlate with lower congestion, and this is indeed the case for a majority of client IP addresses contained in the M-Lab data. However, for some IP addresses, we observe strong positive correlations (\ie $\rho>0$). Our hypothesis for this unexpected phenomenon is that when multiple houses of an ISP network are sharing an IP address; the speed measurements can vary in a wide range depending on broadband capacity of individual households, whereas congestion counts would have smaller variations reflecting the condition of the network. Thus, having mixed measurements (speed and congestion-count)  from multiple households will likely result in imbalanced data pairs causing an unexpected positive correlation between speed and congestion-count.

To better visualize our discussion and hypothesis, we present in Fig.~\ref{fig:Corr} samples of the correlation between the download-speed and congestion-count observed over a 12-month period from two IP addresses. In each plot, the normalized density distribution of download-speed measurements is depicted by solid black lines. We overlay it by scatter plot of download-speed (x-axis) and its corresponding congestion-counts (y-axis), shown by square/circle markers. Note that for a given IP address, we unit-scale (normalize) measured download-speed and congestion-count separately by dividing each data point by corresponding maximum value (\ie $X_i/X_{max}$ and $Y_i/Y_{max}$; where [$X_i$, $Y_i$] is the pair of download-speed and congestion-count for a client IP).  
In our plots, the scaled value of congestion count for each test-point is proportional to the size of the corresponding marker, tiered in two colors -- low/medium (\ie $<0.5$) congestion counts are in green, and high/very-high (\ie $\geq0.5$)  congestion counts are in red. 

Fig. \ref{fig:NegCorrAgg} shows a negative correlation (\ie $\rho=-0.83$) for 458 test-points obtained from an IP address served by Cox ISP in the US -- smaller green squares are mainly skewed towards the bottom right of the plot (low congestion and high speed values), and larger red circles are grouped at the top left region of the plot (\ie high congestion and low speed values). On the other hand, Fig. \ref{fig:PosCorrAgg} shows a positive correlation ($\rho=0.69$) for 896 test-points from City ISP in the US --  smaller green squares are mainly spread from left to middle bottom of the plot (low congestion and low/medium speed values), and larger red circles are clustered at top middle of the plot (high congestion and medium/high speed values). 



\subsection{Large-Scale Consistency Validation}
We now go back to our M-Lab data to analyze the $\rho$ parameter across various ISPs of different size as well as across months checking whether a consistent pattern of correlation is observed.

\subsubsection{Across ISPs}
Large ISPs such as AT\&T and Comcsat in the US with a wealth of public IP addresses (\ie 91 million and 51 million) 
Smaller ISPs who own smaller pool of IPv4 addresses (\eg class C blocks) are more likely forced to employ NAT (or dynamic lease in the best case) for better management of their limited address resources. On the other hand, larger ISPs who were assigned class A address blocks would have discretion to statically allocate one public IP address to each of their clients.   


\begin{table*}[!t]
	\centering
	\caption{ISP pairs with common support in AU and US (refined dataset).}
	\label{tab:ISPpairsRefined}
	\begin{adjustbox}{max width=0.995\textwidth}
		\renewcommand{\arraystretch}{1.3}      	
		\begin{tabular}{|l|c|c|c|c|c|c|c|c|c|c|c|c|c|c|c|c|c|c|c|c|c|}
			\hline 
			\textbf{ISP pair ID} & \textbf{1} & \textbf{2} & \textbf{3} & \textbf{4} & \textbf{5} & \textbf{6} & \textbf{7} & \textbf{8} & \textbf{9} & \textbf{10} & \textbf{11} & \textbf{12} & \textbf{13} & \textbf{14} & \textbf{15} & \textbf{16} & \textbf{17} & \textbf{18} & \textbf{19} & \textbf{20} & \textbf{21}\tabularnewline			
			\hline 
			\textbf{ISP1 vs ISP2} &  \rotatebox{90}{AT\&T vs. Cox} &  \rotatebox{90}{Virgin vs. BT} &  \rotatebox{90}{Comcast vs. AT\&T} &  \rotatebox{90}{Comcast vs. Verizon} &  \rotatebox{90}{Virgin vs. BT} &  \rotatebox{90}{Comcast vs. Verizon} & \rotatebox{90}{Comcast vs. Cox} &  \rotatebox{90}{Virgin vs. BT} &  \rotatebox{90}{Comcast vs. Verizon} &  \rotatebox{90}{Virgin vs. Sky} &  \rotatebox{90}{Telstra vs. Optus} &  \rotatebox{90}{Virgin vs. Sky} &  \rotatebox{90}{Virgin vs. Sky} &  \rotatebox{90}{Virgin vs. Sky} &  \rotatebox{90}{Telstra vs. Optus} &  \rotatebox{90}{Virgin vs. BT} &  \rotatebox{90}{Virgin vs. Sky} &  \rotatebox{90}{Virgin vs. Sky} &  \rotatebox{90}{AT\&T vs. Verizon} &  \rotatebox{90}{Comcast vs. AT\&T} &  \rotatebox{90}{Comcast vs. AT\&T}\tabularnewline
			\hline 
			\textbf{Speed-tier} & {[}50, 75{]} & {[}50, 75{]} & {[}50, 75{]} & {[}75, 100{]} & {[}30, 50{]} & {[}20, 25{]} & {[}50, 75{]} & {[}30, 50{]} & {[}25, 30{]} & {[}30, 50{]} & {[}25, 30{]} & {[}20, 25{]} & {[}25, 30{]} & {[}25, 30{]} & {[}50, 75{]} & {[}12, 20{]} & {[}30, 50{]} & {[}12, 20{]} & {[}75, 100{]} & {[}30, 50{]} & {[}20, 25{]}\tabularnewline
			\hline 
			\textbf{Year} & 2016 & 2015 & 2015 & 2015 & 2015 & 2015 & 2015 & 2016 & 2015 & 2016 & 2016 & 2015 & 2015 & 2016 & 2016 & 2015 & 2015 & 2016 & 2016 & 2015 & 2015\tabularnewline
			\hline\hline 			
			
			\textbf{ISP pair ID} & \textbf{22} & \textbf{23} & \textbf{24} & \textbf{25} & \textbf{26} & \textbf{27} & \textbf{28} & \textbf{29} & \textbf{30} & \textbf{31} & \textbf{32} & \textbf{33} & \textbf{34} & \textbf{35} & \textbf{36} & \textbf{37} & \textbf{38} & \textbf{39} & \textbf{40} & \textbf{41} & \textbf{42}\tabularnewline
			\hline 
			\textbf{ISP1 vs ISP2} & \rotatebox{90}{Virgin vs. TalkTalk} & \rotatebox{90}{Comcast vs. Cox} & \rotatebox{90}{AT\&T vs. Cox} & \rotatebox{90}{Virgin vs. Sky} & \rotatebox{90}{Comcast vs. AT\&T} & \rotatebox{90}{Comcast vs. AT\&T} & \rotatebox{90}{TPG vs. Optus} & \rotatebox{90}{Comcast vs. AT\&T} & \rotatebox{90}{TPG vs. Telstra} & \rotatebox{90}{Comcast vs. Verizon} & \rotatebox{90}{Comcast vs. Cox} & \rotatebox{90}{Comcast vs. Cox} & \rotatebox{90}{AT\&T vs. Cox} & \rotatebox{90}{Comcast vs. Cox} & \rotatebox{90}{Comcast vs. Verizon} & \rotatebox{90}{Comcast vs. Cox} & \rotatebox{90}{Comcast vs. Cox} & \rotatebox{90}{Telstra vs. iiNet} & \rotatebox{90}{Comcast vs. AT\&T} & \rotatebox{90}{TPG vs. Telstra} & \rotatebox{90}{Telstra vs. iiNet}\tabularnewline
			\hline 
			\textbf{Speed-tier} & {[}30, 50{]} & {[}30, 50{]} & {[}8, 12{]} & {[}12, 20{]} & {[}25, 30{]} & {[}8, 12{]} & {[}12, 20{]} & {[}0, 8{]} & {[}12, 20{]} & {[}50, 75{]} & {[}75, 100{]} & {[}12, 20{]} & {[}12, 20{]} & {[}8, 12{]} & {[}0, 8{]} & {[}0, 8{]} & {[}25, 30{]} & {[}12, 20{]} & {[}12, 20{]} & {[}8, 12{]} & {[}0, 8{]}\tabularnewline
			\hline 
			\textbf{Year} & 2016 & 2015 & 2016 & 2015 & 2015 & 2015 & 2015 & 2015 & 2015 & 2015 & 2015 & 2015 & 2016 & 2015 & 2015 & 2015 & 2015 & 2016 & 2015 & 2015 & 2016\tabularnewline
			\hline \hline
			
			\textbf{ISP pair ID} & \textbf{43} & \textbf{44} & \textbf{45} & \textbf{46} & \textbf{47} & \textbf{48} & \textbf{49} & \textbf{50} & \textbf{51} & \textbf{52} & \textbf{53} & \textbf{54} & \textbf{55} & \textbf{56} & \textbf{57} & \textbf{58} & \textbf{59} & \textbf{60} & \textbf{61} & \textbf{62} & \textbf{63}\tabularnewline
			\hline 
			\textbf{ISP1 vs ISP2} & \rotatebox{90}{Telstra vs. iiNet} & \rotatebox{90}{Comcast vs. Verizon} & \rotatebox{90}{Comcast vs. Verizon} & \rotatebox{90}{AT\&T vs. Verizon} & \rotatebox{90}{Comcast vs. Verizon} & \rotatebox{90}{Telstra vs. TPG} & \rotatebox{90}{Telstra vs. Optus} & \rotatebox{90}{Telstra vs. TPG} & \rotatebox{90}{Comcast vs. Cox} & \rotatebox{90}{Telstra vs. Optus} & \rotatebox{90}{Telstra vs. Optus} & \rotatebox{90}{Telstra vs. TPG} & \rotatebox{90}{Telstra vs. iiNet} & \rotatebox{90}{Virgin vs. Sky} & \rotatebox{90}{Virgin vs. Sky} & \rotatebox{90}{Virgin vs. BT} & \rotatebox{90}{TPG vs. Telstra} & \rotatebox{90}{TPG vs. iiNet} & \rotatebox{90}{AT\&T vs. Comcast} & \rotatebox{90}{Telstra vs. iiNet} & \rotatebox{90}{Telstra vs. Optus}\tabularnewline
			\hline 
			\textbf{Speed-tier} & {[}8, 12{]} & {[}8, 12{]} & {[}30, 50{]} & {[}50, 75{]} & {[}12, 20{]} & {[}8, 12{]} & {[}0, 8{]} & {[}0, 8{]} & {[}20, 25{]} & {[}8, 12{]} & {[}12, 20{]} & {[}12, 20{]} & {[}20, 25{]} & {[}0, 8{]} & {[}0, 8{]} & {[}0, 8{]} & {[}0, 8{]} & {[}0, 8{]} & {[}25, 30{]} & {[}30, 50{]} & {[}30, 50{]}\tabularnewline
			\hline 
			\textbf{Year} & 2016 & 2015 & 2015 & 2016 & 2015 & 2016 & 2016 & 2016 & 2015 & 2016 & 2016 & 2016 & 2016 & 2015 & 2016 & 2015 & 2015 & 2015 & 2016 & 2016 & 2016\tabularnewline
			\hline\hline
			
			\textbf{ISP pair ID} & \textbf{64} & \textbf{65} & \textbf{66} & \textbf{67} & \textbf{68} & \textbf{69} & \textbf{70} & \textbf{71} & \textbf{72} & \textbf{73} & \textbf{74} & \textbf{75} & \textbf{76} & \textbf{77} & \textbf{78} & \textbf{79} & \textbf{80} & \multicolumn{3}{l}{\textbf{81}}   & \tabularnewline
			\hline 
			\textbf{ISP1 vs ISP2} & \rotatebox{90}{AT\&T vs. Cox} & \rotatebox{90}{Telstra vs. Optus} & \rotatebox{90}{AT\&T vs. Verizon} & \rotatebox{90}{AT\&T vs. Cox} & \rotatebox{90}{AT\&T vs. Comcast} & \rotatebox{90}{AT\&T vs. Comcast} & \rotatebox{90}{AT\&T vs. Comcast} & \rotatebox{90}{AT\&T vs. Verizon} & \rotatebox{90}{AT\&T vs. Verizon} & \rotatebox{90}{AT\&T vs. Comcast} & \rotatebox{90}{AT\&T vs. Verizon} & \rotatebox{90}{Comcast vs. Cox} & \rotatebox{90}{AT\&T vs. Verizon} & \rotatebox{90}{AT\&T vs. Cox} & \rotatebox{90}{AT\&T vs. Cox} & \rotatebox{90}{AT\&T vs. Comcast} & \rotatebox{90}{AT\&T vs. Comcast} &  \multicolumn{3}{l}{\rotatebox{90}{AT\&T vs. Comcast}}  & \tabularnewline
			\hline 
			\textbf{Speed-tier} & {[}75, 100{]} & {[}20, 25{]} & {[}0, 8{]} & {[}0, 8{]} & {[}8, 12{]} & {[}0, 8{]} & {[}75, 100{]} & {[}8, 12{]} & {[}25, 30{]} & {[}12, 20{]} & {[}20, 25{]} & {[}100, 1000{]} & {[}12, 20{]} & {[}25, 30{]} & {[}30, 50{]} & {[}20, 25{]} & {[}50, 75{]} &  \multicolumn{3}{l}{{[}30, 50{]}}  & \tabularnewline
			\hline 
			\textbf{Year} & 2016 & 2016 & 2016 & 2016 & 2016 & 2016 & 2016 & 2016 & 2016 & 2016 & 2016 & 2015 & 2016 & 2016 & 2016 & 2016 & 2016 & \multicolumn{3}{l}{2016}  & \tabularnewline
			\hline 
		\end{tabular}		
	\end{adjustbox}
\end{table*}

We, therefore, start examining the aggregate $\rho$ parameter for each ISP  in AU and US. We select two of large and two of small ISPs from each country for comparison: in Australia, Telstra and Optus as large providers, and Harbour and CEnet as small providers; in the US, Comcast and AT\&T as large providers, and Hurricane and Lightower as small providers. We present in Fig.~\ref{fig:rhoAUUS} the normalized density distribution of $\rho$ value across unique IP addresses of each ISP.
We find $\smallsim$12K, $\smallsim$4K, 79, and 24 unique addresses from network of Australian ISPs Telstra, Optus, Harbour and CEnet respectively conducting total of $\smallsim$453K, $\smallsim$182K, 5273, and 4638 NDT tests over 12-month period (Aug'16 - Jul'17). Fig.~\ref{fig:rhoAU} shows the $\rho$ distribution for our selected operators in Australia. It is seen that the $\rho$ parameter is predominately negative in large ISPs (shown by solid red lines for Telstra and dashed green lines for Optus in Fig.~\ref{fig:rhoAU}) suggesting that majority of IP addresses present in M-Lab data (from these two large ISPs) are consistently assigned to single households. Moreover, the $\rho$ distribution is fairly biased towards positive values for smaller ISPs -- average $\rho= 0.31$ for Harbour (its distribution is shown by dotted blue lines) and average $\rho= 0.58$ for CEnet (its distribution is shown by dashed-dotted purple lines) in Fig.~\ref{fig:rhoAU}, meaning that IP addresses are mainly shared by multiple households of varied broadband capacity.

\begin{figure}[!t]
	\centering
	\includegraphics[width=0.45\textwidth,height=0.23\textheight]{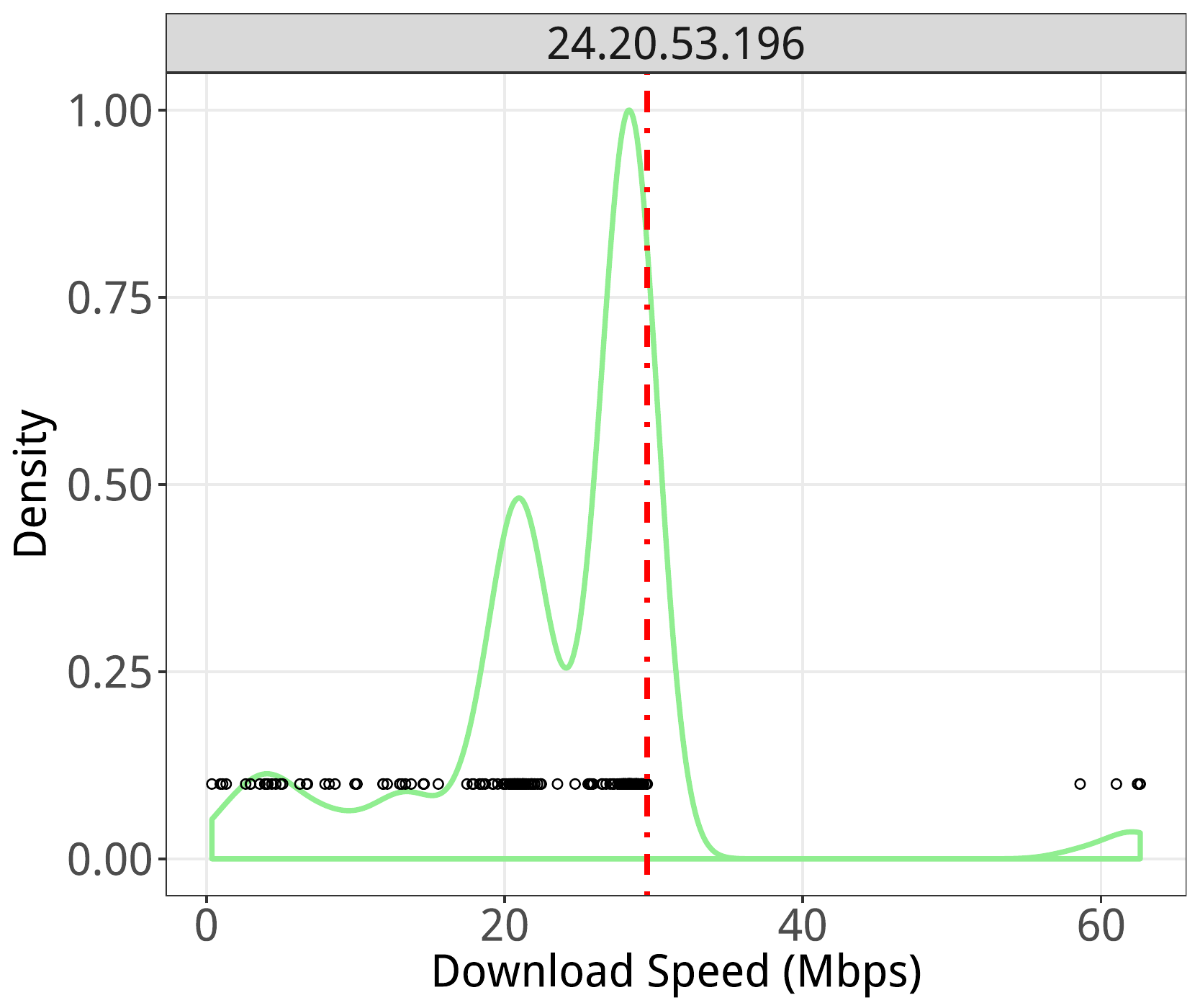}
	\vspace{-3mm}
	\caption{Outliers in speed measurements.}
	\vspace{-4mm}
	\label{fig:outliers}
\end{figure}

Similarly, we observe an aggregate negative correlation values for large ISPs in the US along with neutral/positive correlation for smaller ISPs, as shown in Fig.~\ref{fig:rhoUS}. For our US selected ISPs Comcast, AT\&T, Hurricane, and Lightower, we have $\smallsim$4.3m, $\smallsim$2.8m, $\smallsim$46K, and $\smallsim$14K NDT test-points respectively indexed by $\smallsim$98K, $\smallsim$53K, 424, and 176 unique addresses. The average $\rho$ for large operators Comcast and AT\&T is  $-0.39$ and $-0.43$ respectively, whereas smaller operators Hurricane and Lightower exhibit positive average correlation of $0.21$ and $0.10$ respectively.

We see on average a negative correlation between measured download-speed and congestion-count across large network operators (with large pool of IP addresses), and positive correlation values across small network operators (with small pool of IP addresses) in two countries Australia and US.

\begin{figure}[!t]
	\centering
	\includegraphics[width=0.5\textwidth]{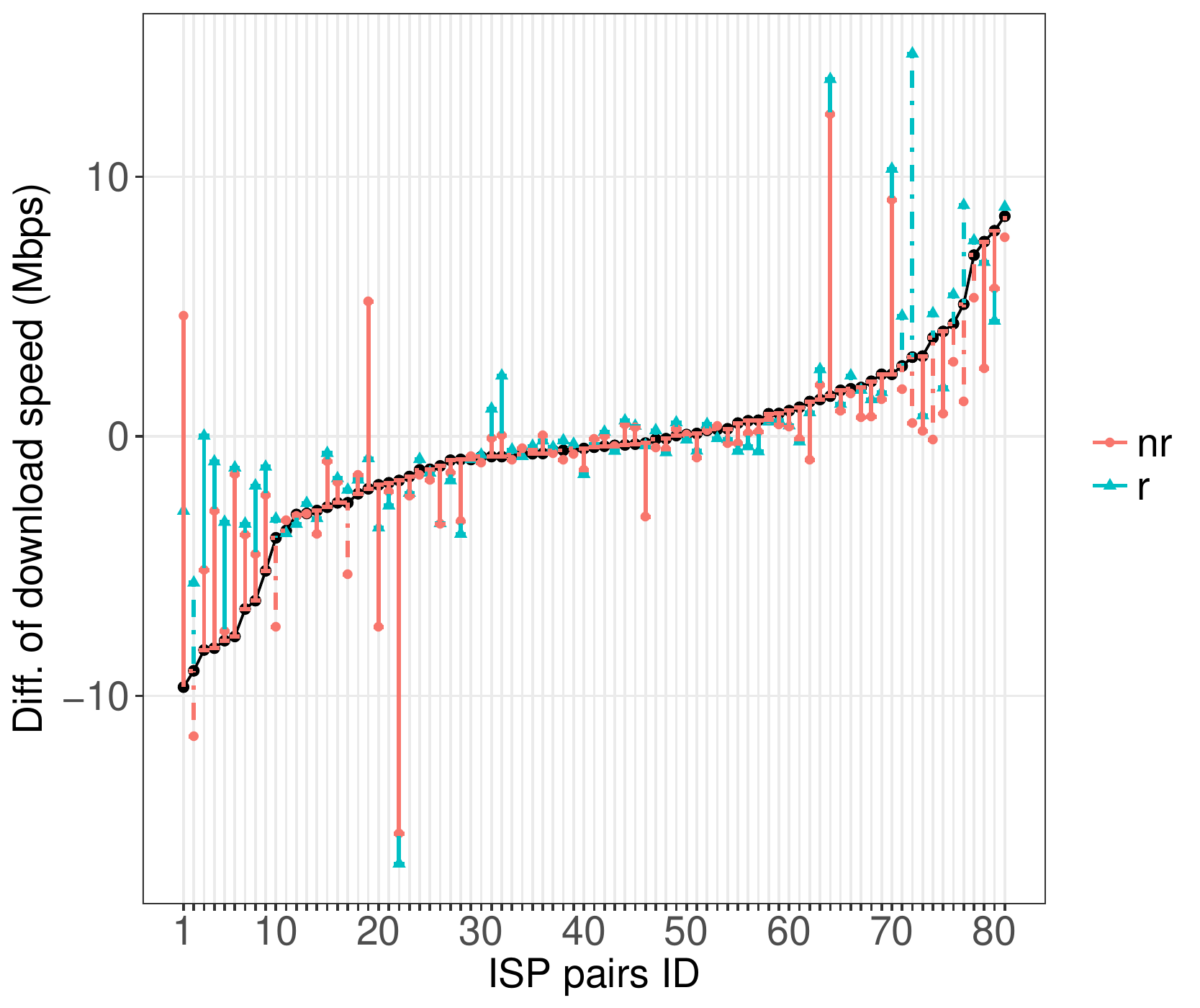}
	\vspace{-6mm}
	\caption{Comparing ISP speeds for 81 pairs in AU and US (refined dataset).}
	\vspace{-4mm}
	\label{fig:ISPpairsRefined}
\end{figure}

\subsubsection{Across Months}
We now track the correlation value within a network operator across various months to check whether change of network conditions would affect the $\rho$ value. This verifies the validity of our hypothesis over time. We, therefore, compute the $\rho$ value for a given IP address on a monthly basis using data points observed within a month, \eg April 2017. 

In Fig.~\ref{fig:Corr}, we saw speed, congestion and the $\rho$ value computed on aggregate data of 12-month period for one sample of IP address in each network (large and small separately). We visualize in Fig.~\ref{fig:CorrMnt}  the monthly data along with corresponding $\rho$ values for the same IP addresses and their respective networks. 
We observe an strong negative correlation for data of address {\fontsize{10}{48}\usefont{OT1}{lmtt}{b}{n}\noindent 98.174.39.22}  from Cox (one of the top ten large ISPs in the US) consistent across four months in 2017, as shown in Fig.~\ref{fig:NegCorrMnt}. Individual monthly speed density curves (narrow single hump) and congestion clusters are fairly similar to the plot in Fig.~\ref{fig:NegCorrAgg}, and the $\rho$ value is $-0.90$, $-0.88$, $-0.82$, and $-0.76$ for successive months April, May, June, and July respectively.

Considering the IP address from a smaller operator in Fig.~\ref{fig:PosCorrMnt}, a strong positive correlation is observed consentingly across four successive months in 2016-2017. In each plot, download-speed density curve depicts two humps and congestion markers are aligned (green squares on the left and red circles on the right) in opposite direction to that which is expected, just similar to aggregate performance measurements in Fig.~\ref{fig:PosCorrAgg}. We again see a strong positive $\rho$ values of $0.61$, $0.62$, $0.69$, and $0.62$ respectively for November and December in 2016, and January and February in 2017. 

Our analysis of M-Lab data across various network operators and across various months validates that our hypothesis holds true.

\subsection{Estimating Household Speed-Tier}\label{sec:insights} 

We now filter measurements corresponding to those IP addresses that exhibit positive correlation between their download-speed and congestion-count (\ie $\rho > 0$). We note that a large fraction of IP addresses from small ISPs are filtered due to positive $\rho$ value. For example, no data from CEnet (in AU) is considered as single house \cite{arXiv2019}.  


After removing data of multiple households, we estimate the \textit{speed-tier} as a proxy for broadband capacity of each house. 
We term this maximum possible speed available to the household as its ``speed-tier''. 
As far as maximum download speed is concerned, in some cases we observe very large values in measurements which are more likely to be outliers. Fig.~\ref{fig:outliers} exemplifies measured download speed from a sample household. We use green solid lines to show the density distribution of speed overlayed by black circles stacked along the x-axis representing actual data points. We can see that there are several outliers observed around 60 Mbps in   Fig.~\ref{fig:outliers} while the rest of measurements fall under 30 Mbps -- the maximum speed value seems to be about half of outliers value. The dashed vertical red line depicts the cut-off point to filter outliers.

In order detect and exclude outliers data in our study, we employ the standard modified Thompson Tau technique \cite{Tau} to statistically determine rejection zone. This method eliminates outliers more than two standard deviations from the mean value. After filtering outliers from our dataset, we pick the maximum value of remaining data points as the estimated speed-tier of corresponding house (\ie IP address) \cite{arXiv2019}. 

\subsection{Multi-Variate Matching on Refined Dataset} \label{sec:matchingRefined} 
Lastly, we apply the multi-variate matching technique on the refined dataset (\ie removing data-points of households with positive $\rho$ followed by eliminating outliers) to re-evaluate our comparison of ISP performance.
Similar to previous section, we begin by sorting ISP pairs in ascending order of their average difference of download speed. Note that we now have 81 ISP pairs (\ie 2 pairs less than the original dataset in the previous section) with sufficient common support -- new pairs are listed in Table~\ref{tab:ISPpairsRefined}. Our matching results are shown in Fig.~\ref{fig:ISPpairsRefined}. We observe that for majority of ISP pairs, matching estimated differences are closer to zero compared to the naive average -- though we see a few exceptions (\eg pair IDs 22, 64, 72). This reiterates our view that when ISPs are compared fairly by adjusting for test conditions, they are not so different.

\section{Conclusion} \label{sec:concl}
This paper is a first step towards a fair comparison of speed performance across broadband ISPs, by applying emerging causal inference techniques widely used in medicine to the large volume of measurement data from M-Lab. We first built a tool to pre-process and visualize M-Lab data, giving preliminary insights into the factors affecting speed performance. We then demonstrated that test attributes such as access speed-tier, host TCP window size, and server distance vary in distribution across ISPs, and further that these attributes affect measurement outcomes. We then applied multi-variate matching to reduce the confounding bias, and our fair comparison between pairs revealed that the difference between ISPs is much lower than what naive aggregates may suggest. 
Our future work will expand this study by estimating the comparative performance of ISPs for individual customers rather than just aggregates. This will be achieved using more sophisticated methods, such as machine learning based Targeted Maximum Likelihood (TML) algorithms,  which can deal with both confounding as well as differential causal effects.


\bibliographystyle{IEEEtran}
\bibliography{bibMlabPerf}

\begin{IEEEbiography}[{\includegraphics[width=1in,height=1.25in,clip,keepaspectratio]{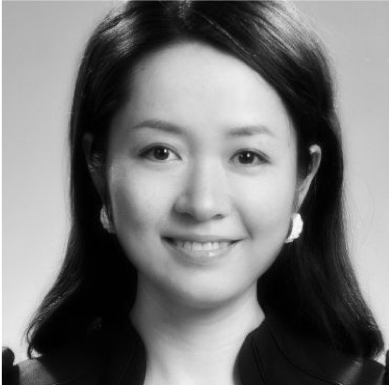}}]{Xiaohong~Deng}
	received her B.Sc. and M.Sc. degrees of Computer Science from Chongqing University of Posts and Telecommunications in 2004 and Beijing University of Posts and Communications in 2007 respectively, and  her Ph.D. in Electrical Engineering and Telecommunications from the University of New South Wales in Sydney, Australia in 2019. She served France Telecom from 2008 to 2013 as a Network Architect and Project Lead. Her research interests include broadband networks and big data analytics of network performance data.
\end{IEEEbiography}

\begin{IEEEbiography}[{\includegraphics[width=1in,height=1.25in,clip,keepaspectratio]{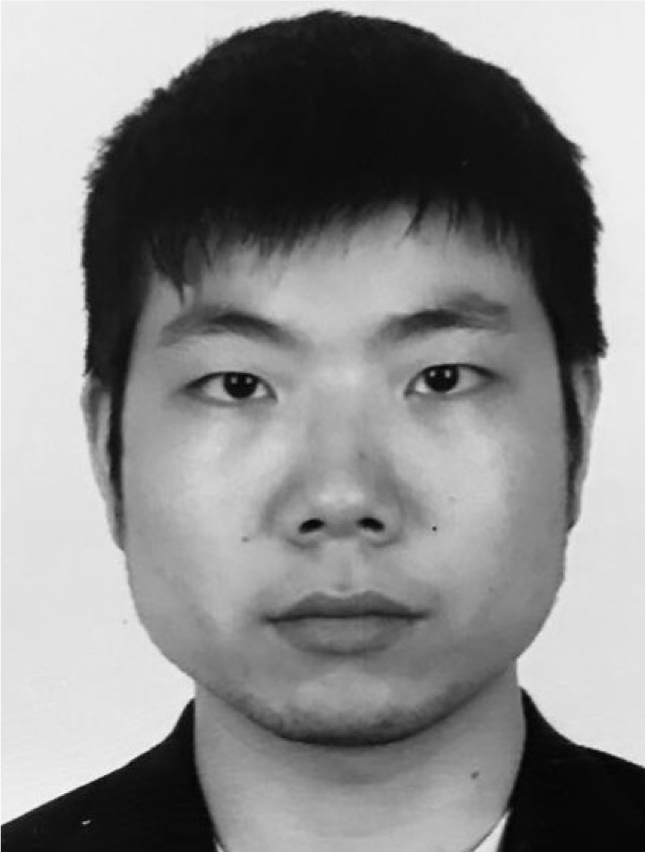}}]{Yun~Feng}
	received his B.Sc. and M.Sc. degrees of Telecommunication from Xidian University in China and the University of New South Wales in Sydney in 2014 and 2018 respectively. He was a research assistant in the School of Electrical Engineering and Telecommunications at the University of New South Wales. He is currently working at Shanghai Huawei Technologies. His research interests include big data, machine learning and application of embedded system. 
\end{IEEEbiography}

\begin{IEEEbiography}[{\includegraphics[width=1in,height=1.25in,clip,keepaspectratio]{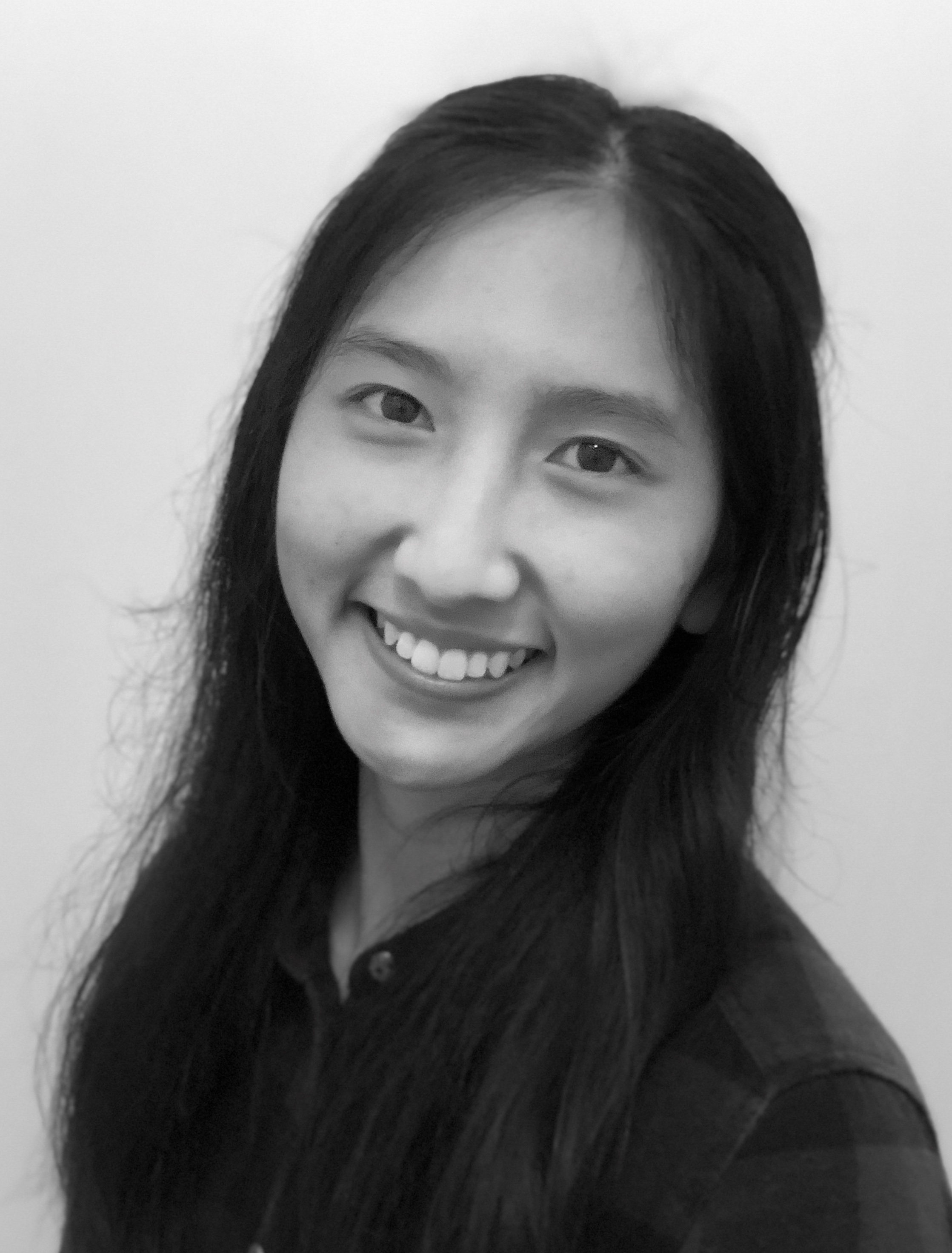}}]{Thanchanok~Sutjarittham}
	is currently pursuing her Ph.D. in Electrical Engineering and Telecommunications at the University of New South Wales (UNSW Sydney), where she has also received her B.Eng. in Electrical Engineering and Telecommunications in 2016. Her primary research interests include Internet of Things, sensors  data analytics, and applied machine learning.
\end{IEEEbiography}

\begin{IEEEbiography}[{\includegraphics[width=1in,height=1.25in,clip,keepaspectratio]{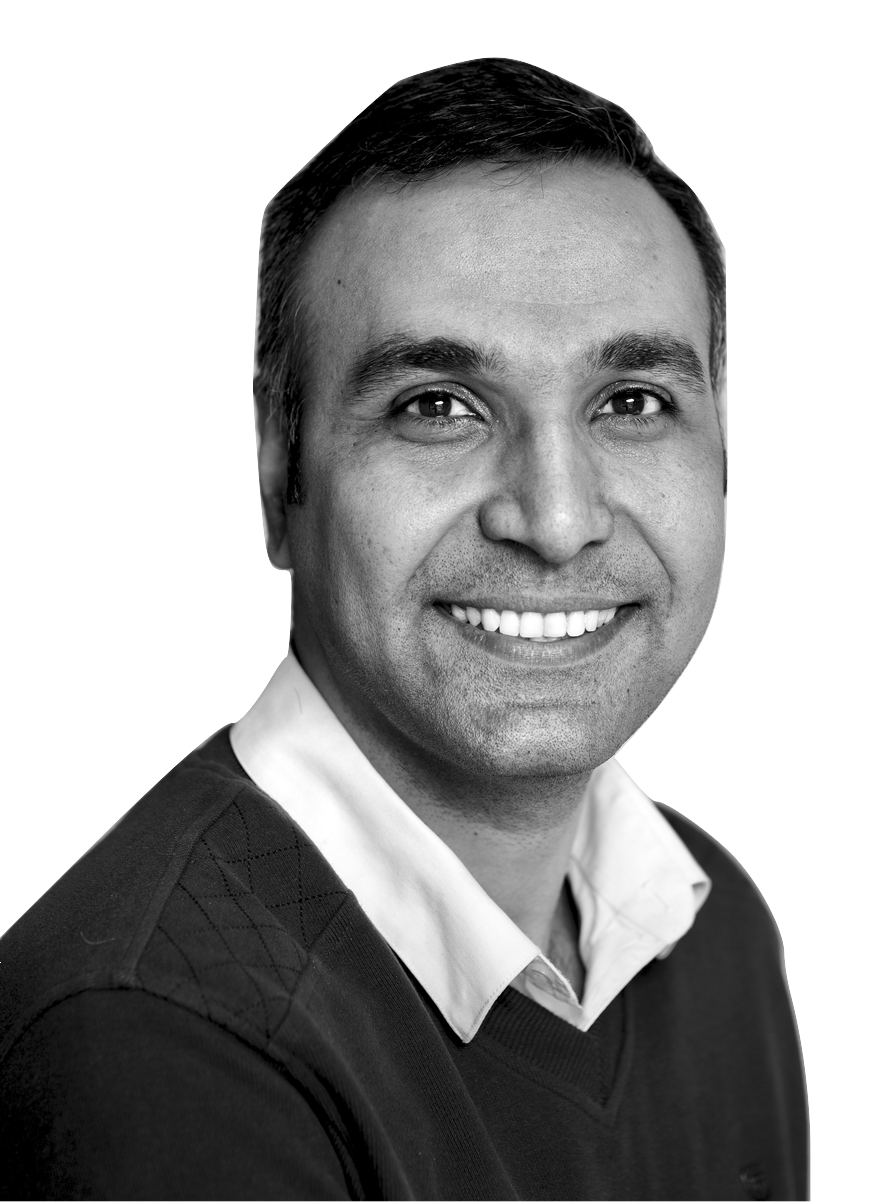}}]{Hassan~Habibi~Gharakheili}
	received his B.Sc. and M.Sc. degrees of Electrical Engineering from the Sharif University of Technology in Tehran, Iran in 2001 and 2004 respectively, and his Ph.D. in Electrical Engineering and Telecommunications from the University of New South Wales in Sydney, Australia in 2015. He is currently a Senior Lecturer at the University of New South Wales in Sydney, Australia. His current research interests include programmable networks, learning-based networked systems, and data analytics in computer systems.
\end{IEEEbiography}

\begin{IEEEbiography}[{\includegraphics[width=1in,height=1.25in,clip,keepaspectratio]{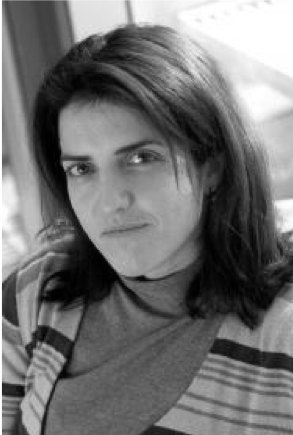}}]{Blanca Gallego Luxan}
	received the B.S. degree from the Universidad Autónoma Metropolitana, and the Ph.D. degree from the University of California, Los Angeles. She is currently an Associate Professor at the Centre for Big Data Research in Health, UNSW. She has extensive international research experience in data analysis and computational modeling and has made significant and innovative contributions to the design, analysis, and development of models derived from complex empirical data for a wide range of applications, such as patient safety, biosurveillance, corporate sustainability reporting, ecological footprint analysis, and climate variability.
\end{IEEEbiography}

\begin{IEEEbiography}[{\includegraphics[width=1in,height=1.25in,clip,keepaspectratio]{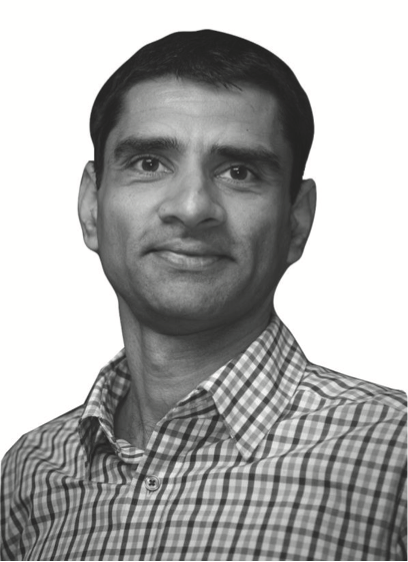}}]{Vijay Sivaraman}
	received his B. Tech. from the Indian Institute of Technology in Delhi, India, in 1994, his M.S. from North Carolina State University in 1996, and his Ph.D. from the University of California at Los Angeles in 2000. He has worked at Bell-Labs as a student Fellow, in a silicon valley start-up manufacturing optical switch-routers, and as a Senior Research Engineer at the CSIRO in Australia. He is now a Professor at the University of New South Wales in Sydney, Australia. His research interests include Software Defined Networking, network architectures, and cyber-security particularly for IoT networks.
\end{IEEEbiography}
 
\end{document}